\documentclass[11pt]{article}
\usepackage[utf8]{inputenc}
\usepackage{jheppub}
\usepackage{graphicx}
\usepackage{caption}
\usepackage{subcaption}
\usepackage{hyperref}
\usepackage[T1]{fontenc}
\usepackage{float}

\newcommand{\be}{\begin{equation}}
\newcommand{\ee}{\end{equation}}
\newcommand{\bea}{\begin{eqnarray}}
\newcommand{\eea}{\end{eqnarray}}

\newcommand\underrel[3][]{\mathrel{\mathop{#3}\limits_{%
      \ifx c#1\relax\mathclap{#2}\else#2\fi}}}
\title{Improving the five-point bootstrap}

\author{David Poland,$^{a}$ Valentina Prilepina,$^{b}$ Petar Tadi\' c$^{a}$}

\affiliation{$^{a}$ Department of Physics, Yale University, New Haven, CT 06520, USA}
\affiliation{$^{b}$ Perimeter Institute for Theoretical Physics, Waterloo, ON N2L 2Y5, Canada}

\emailAdd{david.poland@yale.edu, valentina.prilepina.1@ulaval.ca, petar.tadic@yale.edu}

\abstract{We present a new algorithm for the numerical evaluation of five-point conformal blocks in $d$-dimensions, greatly improving the efficiency of their computation. To do this we use an appropriate ansatz for the blocks as a series expansion in radial coordinates, derive a set of recursion relations for the unknown coefficients in the ansatz, and evaluate the series using a Pad\' e approximant to accelerate its convergence. We then study the $\langle\sigma\sigma\epsilon\sigma\sigma\rangle$ correlator in the 3d critical Ising model by truncating the operator product expansion (OPE) and only including operators with conformal dimension below a cutoff $\Delta\leqslant \Delta_{\rm cutoff}$. We approximate the contributions of the operators above the cutoff by the corresponding contributions in a suitable disconnected five-point correlator. Using this approach, we compute a number of OPE coefficients with greater accuracy than previous methods.}  

\begin{document}
\maketitle
\flushbottom

\newpage
\section{Introduction}

Theoretical consistency of conformal field theories (CFTs) places non-perturbative bounds on their observables and therefore allows for insights into the dynamics of such theories at strong coupling.  These conditions rely on conformal symmetry as well as the internal symmetries of the theory. The study of these conditions, a.k.a.~the conformal bootstrap, developed following \cite{Ferrara:1973yt, Polyakov:1974gs, Rattazzi:2008pe, El-Showk:2012cjh}, and has led to remarkable progress in understanding the space of conformal field theories and their dynamics. When imposing such consistency conditions, one possible goal is to determine the CFT data of a particular theory. This data consists of its spectrum of primary operators, i.e. their conformal dimensions, along with the three-point correlators of all primary operators, which are fixed by symmetry up to a finite number of unknown coefficients, called operator product expansion (OPE) coefficients. 

In principle, one can access all CFT data by imposing consistency conditions, in particular crossing symmetry, on the four-point correlation functions of arbitrary external operators. In practice, this approach becomes very challenging as soon as some of the external operators in the four-point functions have a non-zero spin (see e.g.~\cite{Iliesiu:2015qra, Iliesiu:2017nrv, Dymarsky:2017xzb, Dymarsky:2017yzx, Karateev:2019pvw, Reehorst:2019pzi, Erramilli:2020rlr, Erramilli:2022kgp, He:2023ewx}).  Instead, by studying correlation functions involving more than four scalar operators, it is possible to access the same CFT data as that accessible by means of the four-point bootstrap with external spinning operators.

Over the past few years, some preliminary steps were made in the attempt to bootstrap higher-point correlation functions, both numerically \cite{Poland:2023vpn} and analytically \cite{Bercini:2020msp, Antunes:2021kmm, Anous:2021caj, Kaviraj:2022wbw, Goncalves:2023oyx, Costa:2023wfz}. An immediate obstruction to bootstrapping higher-point correlators is the lack of ready availability of higher-point conformal blocks. However, recently there has been significant progress in understanding these objects. The conformal block for exchanged scalar operators in five-point correlation functions was first computed in \cite{Rosenhaus:2018zqn}. Following this result, higher-point blocks for exchanged scalars were also computed via holographic methods in \cite{Parikh:2019ygo, Parikh:2019dvm, Hoback:2020pgj, Fortin:2022grf} and by means of dimensional reduction in \cite{Hoback:2020syd}. 
The work~\cite{Pal:2020dqf} established an intriguing relation between higher-point conformal blocks and solutions of a Lauricella system, while a connection to Gaudin models was made in \cite{Buric:2020dyz, Buric:2021ywo, Buric:2021ttm, Buric:2021kgy}. The works~\cite{Skiba:2019cmf, Fortin:2019dnq, Fortin:2019zkm, Fortin:2020ncr, Fortin:2020yjz, Fortin:2020bfq} developed general representations of higher-point conformal blocks by making use of the operator product expansion in embedding space, while~\cite{Rosenhaus:2018zqn, Fortin:2020zxw, Fortin:2023xqq} further developed general representations of one- and two-dimensional higher-point blocks.

Going beyond scalar exchange in higher dimensions, a series expansion for five-point conformal blocks with exchanged spinning operators (and identical external scalars) was computed in \cite{Goncalves:2019znr}. Recursion relations that relate the five-point conformal blocks for exchanged operators of different spin were derived in \cite{Poland:2021xjs}. These allow one to express the five-point block for exchanged operators of arbitrary spin as a finite sum of five-point blocks with exchanged scalar operators. In \cite{Poland:2023vpn} it was further shown that the five-point conformal blocks for exchanged operators of arbitrary spin can be computed by using an expansion in radial coordinates and fixing the unknown coefficients in the ansatz by perturbatively solving two quadratic Casimir differential equations. While this approach was successful overall, it became technically challenging to carry out for exchanged spins $\geq 6$. 

A further obstacle that arises when implementing the five-point numerical bootstrap is due to the fact that the methods used cannot depend on unitarity, as it is not possible to write positivity conditions involving five-point functions.  The method employed in \cite{Poland:2023vpn} was based on truncation techniques originally developed for four-point correlators in \cite{Gliozzi:2013ysa, Gliozzi:2014jsa} and further applied and developed in \cite{Gliozzi:2015qsa, Nakayama:2016cim, Gliozzi:2016cmg, Esterlis:2016psv, Hikami:2017hwv, Hikami:2017sbg, Li:2017agi, Li:2017ukc, Hikami:2018mrf, Leclair:2018trn, Rong:2020gbi, Nakayama:2021zcr, Li:2021uki, Kantor:2021kbx, Kantor:2021jpz,  Laio:2022ayq, Kantor:2022epi, Niarchos:2023lot, Li:2023tic}. To improve the results of \cite{Poland:2023vpn}, it is necessary to consider more exchanged primary contributions that are not truncated. 

In this paper we describe an approach where we approximate the truncated part of the five-point correlator $\langle\sigma\sigma\epsilon\sigma\sigma \rangle$ in the 3d critical Ising model with the corresponding contributions in a disconnected five-point correlator of the form $\langle \sigma \sigma \rangle \langle \epsilon \sigma \sigma \rangle + ({\text {permutations}})$. To run this technique effectively, we also develop a new method for the numerical evaluation of the five-point conformal blocks. The new method is similar to the technique developed for four-point conformal blocks in \cite{Costa:2016xah}; namely, using two quadratic Casimir equations, we write recursion relations that the coefficients in the ansatz for the five-point conformal blocks satisfy and then solve them numerically. To further accelerate the convergence of the blocks and their derivatives, we use a Pad\' e approximant. This is the most efficient method to date for the numerical evaluation of the five-point conformal blocks for exchanged operators of arbitrary spin.

With our disconnected correlator approximation for the truncated part of the $\langle\sigma\sigma\epsilon\sigma\sigma \rangle$ correlator, we observe a shift in the predictions for OPE coefficients compared to what we computed in \cite{Poland:2023vpn}, with the new values given by
\begin{equation}
\begin{split}
&\lambda_{T\epsilon T}^{0}\approx 0.958(7)\,,\\
&\lambda_{T\epsilon C}^{0}\approx 0.48(2)\,,\\
&\lambda_{C\epsilon C}^{4}\approx -0.28(2)\,,
\end{split}
\end{equation} 
where $\epsilon$ is the leading $\mathbb{Z}_{2}$-even scalar, $T$ is the spin-2 stress tensor, and $C$ is the leading spin-4, $\mathbb{Z}_{2}$-even operator. The superscripts in the OPE coefficients denote different constituent tensor structures of the corresponding three-point function. We use the standard box basis of the conformally invariant tensor structures in the three-point functions, defined in \cite{Costa:2011mg}. While the error bars are not rigorous and only estimated, we note that they are significantly smaller than the comparable error bars from~\cite{Poland:2023vpn}.

This paper is structured as follows. In section~\ref{five-point-blocks} we discuss our new approach for the numerical evaluation of five-point conformal blocks with arbitrary spins of the exchanged operators. In section~\ref{four-point-sec} we study the $\langle\sigma\sigma\sigma\sigma\rangle$ correlator in the 3d critical Ising model using the truncation technique and show how a mean-field theory approximation for the operators with conformal dimension above the cutoff $\Delta_{\rm cutoff}$ improves the predictions for the OPE coefficients. In section~\ref{five-point-mft-sec} we study the disconnected five-point correlator $\langle\sigma\sigma\epsilon\sigma\sigma \rangle_{d}$. In section~\ref{five-point-numerics} we analyze the $\langle \sigma\sigma\epsilon\sigma\sigma\rangle$ correlator in the 3d critical Ising model using the truncation technique and approximating the truncated contributions by their counterparts from the disconnected correlator. We discuss our results in section~\ref{disc-sec}.

{\bf Note added:} As our draft was being completed we learned of the work~\cite{Li:2023tic}, which has overlap with our section~\ref{four-point-sec}. The work~\cite{Antunes:2023dlk} also appeared which computes numerical bounds from six-point correlators.

\section{Five-point conformal blocks}\label{five-point-blocks}

In this section we present a new and more efficient method for the numerical evaluation of the five-point conformal blocks.

We start by considering a five-point correlation function of external scalar operators 
$$
\langle\phi_1(x_1)\phi_2(x_2)\phi_3(x_3)\phi_4(x_4)\phi_5(x_5)\rangle\,.
$$
Using the operator product expansion in the (12) and (45) channels, this correlator can be
written as the sum of conformal blocks as
\begin{equation}\label{first-channel}
\begin{split}
&\langle\phi_1(x_1)\phi_2(x_2)\phi_3(x_3)\phi_4(x_4)\phi_5(x_5)\rangle =\\
&\sum_{(\mathcal{O}_{\Delta,l},\mathcal{O'}_{\Delta',l'})}\sum_{n_{IJ}=0}^{{\rm min}(l,l')} \lambda_{\phi_1\phi_2 \mathcal{O}_{\Delta,l}}\lambda_{\phi_4\phi_5 \mathcal{O'}_{\Delta',l'}}\lambda_{\mathcal{O}_{\Delta,l} \phi_3 \mathcal{O'}_{\Delta',l'}}^{n_{IJ}}P(x_i)G^{(n_{IJ})}_{(\Delta, l, \Delta', l')}(u_1',v_1',u_2',v_2',w')\,.
\end{split}
\end{equation}
The conformal blocks $G^{(n_{IJ})}_{(\Delta, l, \Delta', l')}$ incorporate the contributions of two symmetric traceless primary operators $(\mathcal{O}_{\Delta,l},\mathcal{O'}_{\Delta',l'})$ to the five-point correlator, along with their descendants, which show up in the respective operator product expansions $\phi_1 \times \phi_2$ and $\phi_4 \times \phi_5$. Conformal symmetry completely determines the form of these contributions. The sums over $\mathcal{O}_{\Delta,l}$ and $\mathcal{O'}_{\Delta',l'}$ span all the primary operators in the spectrum of the theory.  Since the three-point function $\langle \mathcal{O}_{\Delta,l} \phi_3 \mathcal{O'}_{\Delta',l'} \rangle$ contains multiple independent conformally-invariant tensor structures, the label $n_{IJ}$ is present to enumerate these various structures, with $n_{IJ} = 0,1,\ldots, {\rm min}(l,l')$.

We follow the conventions of \cite{Rosenhaus:2018zqn}. In these conventions the pre-factor function $P(x_i)$ (commonly called the external leg factor) and the cross-ratios are as follows:
\begin{equation}\label{cros-rat-one}
\begin{split}
&P(x_i)=\frac{1}{x_{12}^{\Delta_{1}+\Delta_{2}} x_{34}^{\Delta_{3}} x_{45}^{\Delta_{4}+\Delta_{5}}}\left(\frac{x_{23}}{x_{13}}\right)^{\Delta_{12}}\left(\frac{x_{24}}{x_{23}}\right)^{\Delta_{3}}\left(\frac{x_{35}}{x_{34}}\right)^{\Delta_{45}}\,, \qquad x_{ij}=x_i-x_j\,,\\
&u_1'=\frac{x_{12}^{2}x_{34}^{2}}{x_{13}^{2}x_{24}^{2}}\,,\quad v_1'=\frac{x_{14}^{2}x_{23}^{2}}{x_{13}^{2}x_{24}^{2}}\,,\quad u_2'=\frac{x_{23}^{2}x_{45}^{2}}{x_{24}^{2}x_{35}^{2}}\,,\quad v_2'=\frac{x_{25}^{2}x_{34}^{2}}{x_{24}^{2}x_{35}^{2}}\,,\quad w'=\frac{x_{15}^{2}x_{23}^{2}x_{34}^{2}}{x_{24}^{2}x_{13}^{2}x_{35}^{2}}\,,
\end{split}
\end{equation}
where $\Delta_{i}$ are the conformal dimensions of $\phi_i$ and $\Delta_{ij}=\Delta_{i}-\Delta_{j}$. 

The five-point conformal blocks for exchanged scalar primary operators $(l=l'=n_{IJ}=0)$ were first computed in \cite{Rosenhaus:2018zqn, Parikh:2019dvm, Fortin:2019zkm}. Beyond the scalar case, conformal blocks for exchanged spinning operators can be extracted by means of the recursion relations derived in \cite{Poland:2021xjs}. Using these relations, one can express the five-point conformal block for exchanged spinning operators as a finite sum of conformal blocks for exchanged scalar operators. 

Another method for computing the five-point conformal blocks was presented in \cite{Poland:2023vpn}. There, we used an appropriate ansatz for the five-point conformal blocks as a series expansion in radial coordinates and fixed the coefficients in this ansatz by perturbatively solving two quadratic Casimir differential equations.

In this paper, we improve the method \cite{Poland:2023vpn} so as to efficiently compute blocks for exchanged operators of higher spin. First, we consider the cross-ratios defined in \citep{Buric:2021ywo, Buric:2021kgy} and given by
\begin{equation}
\begin{split}
& u_1'=z_1 \bar{z}_1\,, \qquad v_1'=(1-z_1)(1-\bar{z}_1)\,,\\
& u_2'=z_2 \bar{z}_2\,, \qquad v_2'=(1-z_2)(1-\bar{z}_2)\,,\\
& w'=w(z_1-\bar{z}_1)(z_2-\bar{z}_2)+(1-z_1-z_2)(1-\bar{z}_1-\bar{z}_2)\,.
\end{split}
\end{equation}
Next, we introduce an analog of the radial coordinates~\cite{Hogervorst:2013sma, Costa:2016xah} used for four-point conformal blocks: 
\begin{equation}\label{angluarcoord}
\begin{split}
&z_i=\frac{4 \rho_i}{(1+\rho_i)^2}, \qquad \rho_i = r_i e^{i \theta_i}\,, \qquad \eta_{i}=\cos \theta_i\,, \qquad i=1,2\,, \\
&R=\sqrt{r_1 r_2}\,, \qquad r=\sqrt{\frac{r_1}{r_2}}\,,\qquad \hat{w}=\left(\frac{1}{2}-w\right)\sqrt{(1-\eta_1^2)(1-\eta_2^2)}\,.
\end{split}
\end{equation}
In the coordinates $(R,r,\eta_1,\eta_2,\hat{w})$, the five-point conformal blocks can be written as
\begin{equation}\label{blocks-radial}
\begin{split}
&G^{(n_{IJ})}_{(\Delta,l,\Delta',l')}(R, r, \eta_1, \eta_2, \hat{w})=\\
&\sum_{n=0}^{\infty}R^{\Delta+\Delta'+ n}\sum_{m} \sum_{j_1, j_2} \sum_{k=0}^{{\rm min}(j_1,j_2)} c\left(\frac{n+m}{2},\frac{n-m}{2},j_1,j_2,k\right)r^{\Delta-\Delta'+m} \hat{w}^{k} \eta_{1}^{j_1-k} \eta_{2}^{j_2-k}\,,
\end{split}
\end{equation}
where
\begin{equation}
\begin{split}
&m\in [-n, -n+2, \ldots, n-2, n ]\,,\\
&j_1\in \left[\frac{n+m}{2}+l, \frac{n+m}{2}+l-2, \frac{n+m}{2}+l-4, \ldots , {\rm Mod}\left(\frac{n+m}{2}+l, 2\right) \right]\,,  \\  
&j_2\in \left[\frac{n-m}{2}+l', \frac{n-m}{2}+l'-2, \frac{n-m}{2}+l'-4, \ldots , {\rm Mod}\left(\frac{n-m}{2} +l', 2\right) \right]\,.
\end{split}
\end{equation}
Two obvious advantages of these coordinates as compared to other representations of the blocks are: 1) the simplicity of the angular functions and 2) the property that this expansion only involves only a single infinite sum over the powers of $R$. 

At the zeroth order $(n=m=0)$, we note that there are $1+{\rm min}(l,l')$ independent functions of the angular cross-ratios $\eta_1$, $\eta_2$, and $\hat{w}$ that solve both quadratic Casimir equations. Every such function carries a coefficient $c(0, 0, l, l', k)$, where $k=0,1,\ldots {\rm min}(l,l')$. The rest of the coefficients that enter the ansatz at order zero, namely $c(0,0,j_1,j_2,k)$ for $j_1<l$ and $j_2<l'$, can be determined in terms of $c(0,0,l,l',k)$ upon solving the Casimir equations. The choice of basis for the conformally-invariant tensor structures in $\langle\mathcal{O}_{\Delta,l}\phi_3\mathcal{O'}_{\Delta',l'} \rangle$ correlators then uniquely fixes the coefficients $c(0,0,l,l',k)$. The parameter $n_{IJ}$ therefore appears on the r.h.s. of eq.~(\ref{blocks-radial}) through the coefficients $c(0,0,l,l',k)$. We use the standard box basis, defined in \cite{Costa:2011mg}, where these coefficients are given by 
\begin{equation}
c(0,0,l,l',k)= (-1)^{l+l'+k+n_{IJ}}2^{k+2(\Delta+\Delta')}{n_{IJ} \choose k}\,.
\end{equation}

Let us highlight two apparent differences between the ansatz (\ref{blocks-radial}) and that used in \cite{Poland:2023vpn}. One is in the choice of basis for the angular part, which one is able to choose freely, and the other is in the transformation from the $(r_1, r_2)$ to $(R, r)$ coordinates, which eliminates one of the infinite sums from the ansatz written in \cite{Poland:2023vpn}. Despite this, these two ansatz forms are equivalent.  The form in~(\ref{blocks-radial}) turns out to be more convenient for solving the Casimir equations.

Instead of solving the two quadratic Casimir differential equation order by order in $R$ and $r$, one can follow the logic of  \cite{Costa:2016xah} and derive recursion relations for the $c$-coefficients. See appendix~\ref{recrelapp} for details. Solving these recursion relations numerically is significantly more efficient than solving the Casimir equations perturbatively and allows us to compute $c$-coefficients for significantly larger values of $N_{max}\equiv{\rm Max}(n)$. The algorithm for solving the recursion relations is implemented in an attached {\fontfamily{lmss}\selectfont Mathematica} notebook.
One should note that the number of $c$-coefficients that one needs to compute for a fixed value of $n$ scales as $n^4$. Therefore, the time needed to compute these coefficients in the new algorithm has similar polynomial scaling. 

\begin{figure}[htb]
\centering
\includegraphics[width=0.45\textwidth]{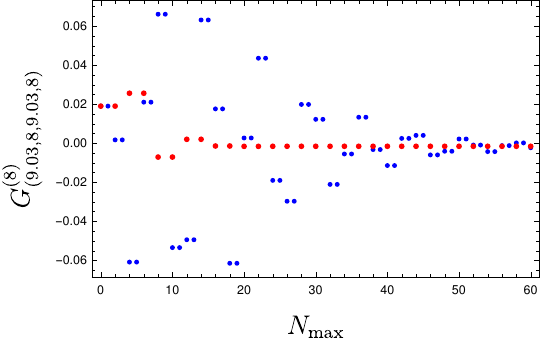}
\hspace{0.05\textwidth}
\includegraphics[width=0.45\textwidth]{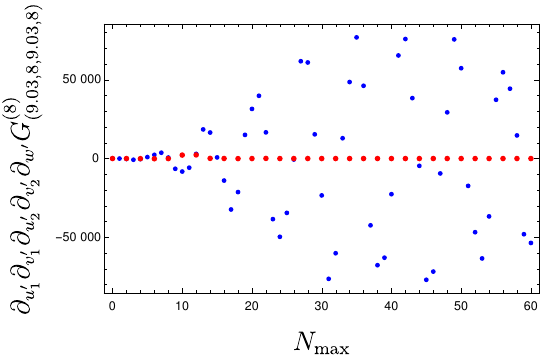}
\caption{{\small  The five-point conformal block and its derivatives with two exchanged operators of spin 8, with the following parameters: $l=l'=n_{IJ}=8$, $\Delta_{12}=\Delta_{45}=0$, $\Delta_{3}=1.413$, $\Delta=\Delta'=9.03$, evaluated at $R=2-\sqrt{3}$, $r=1$, $\eta_1=\eta_2=\hat{w}=0$. The blue points are obtained by evaluating eq.~(\ref{oldsum}), while the red points are obtained using the Pad\' e approximant~(\ref{padesum}). The Pad\' e approximant converges much faster than the original series. Thus, using this method, one can accurately compute five-point conformal blocks (and their derivatives) for exchanged operators with fairly high spin. This particular block cannot be computed with the accuracy shown here by any other available method.}}
\label{blockex}
\end{figure}

To evaluate the blocks, we further utilize a Pad\' e approximant to speed up the convergence of the series expansion. Namely, the conformal block can schematically be represented as
\begin{equation}\label{oldsum}
G\sim (\#) \sum_{n=0}^{N_{\rm max}} a_{n} R^{n}\,,
\end{equation} 
where the coefficients $a_{n}$ depend on all the quantum numbers $(\Delta$, $\Delta'$, $\Delta_{12}$, $\Delta_{3}$, $\Delta_{45}$, $n_{IJ}$, $l$, $l'$), as well as on the remaining cross-ratios $r$, $\eta_1$, $\eta_2$ and $\hat{w}$. We are interested in the dependence of $G$ on $N_{\rm max}$. For each (even) value of $N_{\rm max}$, we can compute a Pad\' e approximant of $G$:
\begin{equation}\label{padesum}
G_{\text{Pad\'e}}\equiv \left[\frac{N_{\rm max}}{2}/\frac{N_{\rm max}}{2}\right]_{G}(R)\,,
\end{equation}
and we observe that the dependence of $G_{\text{Pad\'e}}$ on $N_{\rm max}$ typically converges much faster than $G$. The same is true when we consider the derivatives of the blocks. We show an example of this convergence in fig.~\ref{blockex}. When the expansion is convergent, the series obtained using the Pad\' e approximant is guaranteed to converge to the same value as the original series. Currently, this is the most efficient method for the numerical evaluation of the five-point conformal blocks. 

\section{Four-point $\langle\sigma\sigma\sigma\sigma \rangle$ correlator in the 3d critical Ising model}\label{four-point-sec}

In this section we study the four-point correlator $\langle\sigma\sigma\sigma\sigma\rangle$ in the 3d critical Ising model by truncating the $\sigma \times \sigma$ operator product expansion to include only operators with conformal dimension below a cutoff $\Delta_{\rm cutoff}\approx 5.022$, as in \cite{Gliozzi:2013ysa, Gliozzi:2014jsa}. We will further approximate the contributions of the operators above the cutoff by the corresponding mean-field theory (MFT) contributions. We observe a significant improvement in the prediction of the OPE coefficients of the non-truncated exchanged operators after making this approximation.

The $\langle\sigma\sigma\sigma\sigma\rangle$ correlator in the 3d critical Ising model can be expanded in terms of conformal blocks $g_{\Delta, l}$ in the $(12)(34)$ OPE channel as
\begin{equation}
\langle\sigma(x_1)\sigma(x_2)\sigma(x_3)\sigma(x_4)\rangle = \frac{1}{x_{12}^{2\Delta_\sigma}x_{34}^{2\Delta_{\sigma}}}\sum_{\mathcal{O}_{\Delta,l}} \left(\lambda_{\sigma\sigma \mathcal{O}_{\Delta,l}}\right)^2 g_{\Delta,l}(u,v)\,,
\end{equation}
where $u=\frac{x_{12}^2 x_{34}^2}{x_{13}^2 x_{24}^2}$ and $v=\frac{x_{14}^2 x_{23}^2}{x_{13}^2 x_{24}^2}$. One can also write the expansion in the $(14)(23)$ channel, and by requiring that these two expansions agree one finds the sum rule
\begin{equation}\label{fp-crossrel}
\sum_{\mathcal{O}_{\Delta,l}\neq 1} \mathcal{P}[\mathcal{O}_{\Delta,l}] \mathfrak{f}_{\Delta,l}(u,v) -1=0\,,
\end{equation}
where
\begin{equation}
\begin{split}
&\mathcal{P}[\mathcal{O}_{\Delta,l}]=\left(\lambda_{\sigma\sigma\mathcal{O}_{\Delta,l}}\right)^2\,,\\
&\mathfrak{f}_{\Delta,l}(u,v)=\frac{v^{\Delta_{\sigma}}g_{\Delta,l}(u,v)-u^{\Delta_\sigma}g_{\Delta,l}(v,u)}{u^{\Delta_{\sigma}}-v^{\Delta_{\sigma}}}\,.
\end{split}
\end{equation}
We normalize conformal blocks $g_{\Delta,l}$ as in eq.~(52) in \cite{Poland:2018epd}.
Now, we truncate the sum in eq.~(\ref{fp-crossrel}) by only including the operators in the set $\mathcal{X}=\{\epsilon, \epsilon', T_{\mu\nu}, C_{\mu\nu\rho\sigma}\}$. We fix the conformal dimensions of all exchanged operators using the data \cite{Simmons-Duffin:2016wlq}
\begin{equation}\label{isingconfdim}
\Delta_\epsilon = 1.412625\,, \quad \Delta_{\epsilon'} = 3.82968\,, \quad \Delta_C = 5.022665\,,
\end{equation}
but we treat all OPE coefficients as unknown. We use the following parametrization of the cross-ratios
\begin{equation}
u=\frac{1}{4} \left(a^2-b\right), \qquad v=\left(1-\frac{a}{2}\right)^2-\frac{b}{4}\,.
\end{equation} 
In order to obtain independent constraints we take derivatives of the truncated eq.~(\ref{fp-crossrel}) with respect to the cross-ratios $(a,b)$ and evaluate these at the configuration
\begin{equation}
u=v=\frac{1}{4} \leftrightarrow a=1, \, b=0\,.
\end{equation}
Now, the constraints (which should vanish) are given by
\begin{equation}
e^{\rm 4pt}_{i}(\Delta_{\sigma},\mathcal{P}) = \mathcal{D}_{i}\left(\sum_{\mathcal{O}_{\Delta,l}\in \mathcal{X}} \mathcal{P}[\mathcal{O}_{\Delta,l}] \mathfrak{f}_{\Delta,l}(u,v) -1\right)\Bigg|_{u=v=1/4}\,,
\end{equation}
where $\mathcal{P}$ denotes all (unknown) squares of OPE coefficients and $\mathcal{D}_{i}$ represents derivatives  with respect to $a$ and $b$. 
Here we consider the constraints obtained by taking up to five derivatives with respect to $a$ and $b$. The total number of constraints is then 12. We define $\mathcal{D}_{1}$  as the case where we do not take any derivatives of the crossing relation, i.e. $\mathcal{D}_{1}\equiv 1$. 

Further, we define the cost function $f_{\{r_i\}}^{\rm 4pt}(\Delta_{\sigma},\mathcal{P})$ as
\begin{equation}
f_{\{r_i\}}^{\rm 4pt}(\Delta_{\sigma},\mathcal{P})\equiv \sum_{i=1}^{\mathcal{C}} r_{i} \left(e^{\rm 4pt}_{i}(\Delta_{\sigma},\mathcal{P})\right)^2\,,
\end{equation}
where $r_i$ are (pseudo-)randomly chosen numbers $r_i\in [0,1]$, and $\mathcal{C}$ is the number of constraints we use in the cost function obtained from a given set of derivatives $\mathcal{D}=\{\mathcal{D}_{i}|i=1, 2, \ldots \mathcal{C}\}$. Since we have four unknown OPE coefficients, along with $\Delta_{\sigma}$,  we require at least six constraints in the cost function. In particular, we take $6\leqslant \mathcal{C}\leqslant 9$ in our calculations. We select the sets of constraints such that the value of $\Delta_{\sigma}$ obtained lies closest to the best known numerical value with the smallest standard deviation. Note that we always include the $e_1$ constraint since otherwise the minimization problem would be trivial. The sets $\mathcal{D}$ are given in eq.~(\ref{fourpsetv1}). 

Rather than simply truncating the OPE, let us now consider further approximating all truncated contributions to the crossing relation~(\ref{fp-crossrel}) with the corresponding mean-field theory contributions, which we consider to be double-twist operators $[\sigma\sigma]_{n,l}$ of dimension $2\Delta^{MFT}_{\sigma} + 2n + l$ and spin $l$. In particular we can define the set $\mathcal{X}_{MFT}=\{[\sigma,\sigma]_{0,0}, [\sigma,\sigma]_{1,0}, [\sigma,\sigma]_{0,2}, [\sigma,\sigma]_{0,4}\}$, and then approximate
\begin{equation}
\sum_{\mathcal{O}_{\Delta,l}\notin \mathcal{X}} \mathcal{P}[\mathcal{O}_{\Delta,l}] \mathfrak{f}_{\Delta,l}(u,v)\approx \sum_{[\sigma,\sigma]_{n,l}\notin \mathcal{X}_{MFT}} \left( \lambda_{\sigma\sigma[\sigma,\sigma]_{n,l}}\right)^2 \tilde{\mathfrak{f}}_{2\Delta_{\sigma}^{MFT}+2n+l,l}(u,v)\,,
\end{equation}
where
\begin{equation}
\tilde{\mathfrak{f}}_{2\Delta_{\sigma}^{MFT}+2n+l,l}(u,v)=\frac{v^{\Delta_{\sigma}^{MFT}}g_{\Delta,l}(u,v)-u^{\Delta_\sigma^{MFT}}g_{\Delta,l}(v,u)}{u^{\Delta_{\sigma}^{MFT}}-v^{\Delta_{\sigma}^{MFT}}}\,,
\end{equation}
\begin{equation}\label{mftopecoef}
\lambda_{\sigma \sigma [\sigma,\sigma]_{n,l}}=\frac{\left(1+(-1)^l\right)^{1/2} 2^{l/2} \left(\Delta_{\sigma}^{MFT} -\frac{1}{2}\right)_n (\Delta_{\sigma}^{MFT} )_{l+n}}{\sqrt{l! n! \left(l+\frac{3}{2}\right)_n (n+2 \Delta_{\sigma}^{MFT} -2)_n \left(l+n+2 \Delta_{\sigma}^{MFT} -\frac{3}{2}\right)_n (l+2 n+2 \Delta_{\sigma}^{MFT} -1)_l}}\,,
\end{equation}
and we fix $\Delta_{\sigma}^{MFT}=0.5181489$ to its known numerical value. Note that the scaling dimension $\Delta_{\sigma}^{MFT}$ is a free parameter of the mean-field theory that we set to be equal to the best known value of the conformal dimension of the $\sigma$-operator in the 3d Ising model.

After making this approximation of the truncated spectrum, the constraints are simply given by
\begin{equation}
\begin{split}
&e^{\rm 4pt}_{i}(\Delta_{\sigma},\mathcal{P}) =\\
& \mathcal{D}_{i}\left(\sum_{\mathcal{O}_{\Delta,l}\in \mathcal{X}} \mathcal{P}[\mathcal{O}_{\Delta,l}] \mathfrak{f}_{\Delta,l}(u,v) -\sum_{[\sigma,\sigma]_{n,l}\in \mathcal{X}_{MFT}} \left(\lambda_{\sigma\sigma[\sigma,\sigma]_{n,l}}\right)^2 \tilde{\mathfrak{f}}_{2\Delta_{\sigma}^{MFT}+2n+l,l}(u,v)\right)\Bigg|_{u=v=1/4}\,.
\end{split}
\end{equation}
Even though we fixed $\Delta_{\sigma}^{MFT}$ to be the known value, we still treat $\Delta_{\sigma}$ as a free parameter that we later fix, since these are, in principle, independent parameters in different theories.
Again, for each value of $\mathcal{C}$ in range $6\leqslant \mathcal{C}\leqslant 9$ we choose the set of constraints that give $\Delta_{\sigma}$ closest to its known value with the smallest standard deviation. The sets of derivatives $\mathcal{D}$ that we use are given in eq.~(\ref{fourpsetsv2}). The results of the calculations are summarized in table~\ref{numrestabfourp}.

\begin{table}[t]
\centering
\begin{tabular}{|l|l|l|l|}
\hline
                                & ${\rm no\, MFT}$ & ${\rm with\, MFT}$ & \cite{Simmons-Duffin:2016wlq} \\ \hline
$\Delta_{\sigma}$               & 0.514(5)         & 0.5182(4)          & 0.5181489(10)                  \\ \hline
$\mathcal{P}[\epsilon]$             & 1.15(4)          & 1.106(5)           & 1.106396(9)                     \\
$\mathcal{P}[\epsilon']$            & -0.010(8)        & 0.003(2)           & 0.002810(6)                     \\
$\mathcal{P}[T_{\mu\nu}]$           & 0.33(5)          & 0.422(2)           & 0.425463(1)                     \\
$\mathcal{P}[C_{\mu\nu\rho\sigma}]$ & 0.115(9)         & 0.0768(5)          & 0.0763(1)                     \\ \hline
\end{tabular}
\caption{Numerical data for $\Delta_{\sigma}$ and the unknown OPE coefficients in $\langle\sigma\sigma\sigma\sigma \rangle$ without or with the mean-field theory approximation, compared with the best results for these OPE coefficients from~\cite{Simmons-Duffin:2016wlq}. }\label{numrestabfourp}
\end{table}

We conclude that the mean-field theory approximation for the truncated $\sigma \times \sigma$ operator product expansion works extremely well, motivating us to use a similar approximation when considering the five-point correlators. We note that it is also possible to use the same approximation when one aims to compute the spectrum of the theory as in~\cite{Gliozzi:2013ysa, Gliozzi:2014jsa}, not just the OPE coefficients. With the mean-field theory approximation, one should be able to compute the conformal dimensions of operators below the cutoff with greater accuracy than reported in~\cite{Gliozzi:2013ysa, Gliozzi:2014jsa}.\footnote{This was demonstrated explicitly in the parallel work~\cite{Li:2023tic}.}

\section{Disconnected five-point correlator}\label{five-point-mft-sec}

In this section we study the disconnected five-point correlator $\langle\sigma \sigma \epsilon \sigma \sigma \rangle_{d}$, which will serve as our approximation for the truncated part of the five-point correlator. We define this correlator as the disconnected product of the two-point function $\langle\sigma\sigma \rangle$ and the three-point function $\langle\sigma\sigma\epsilon \rangle$. In this section we expand the disconnected five-point correlator in terms of conformal blocks and compute the OPE coefficients of several contributions. 

More precisely, we define the disconnected five-point correlator as
\begin{equation}
\langle\sigma(x_1) \sigma(x_2) \epsilon(x_3) \sigma(x_4) \sigma(x_5) \rangle_{d} \equiv \langle\sigma(x_1) \sigma(x_2)\rangle \langle \epsilon(x_3) \sigma(x_4) \sigma(x_5) \rangle + ({\rm permutations})\,.
\end{equation}
In terms of cross-ratios, the correlator is explicitly given by
\begin{equation}
\begin{split}
&\langle\sigma(x_1) \sigma(x_2) \epsilon(x_3) \sigma(x_4) \sigma(x_5) \rangle_{d} = \frac{\lambda _{\sigma \sigma \epsilon }}{x_{12}^{2\Delta_{\sigma}}x_{45}^{2\Delta_{\sigma}}x_{34}^{\Delta_{\epsilon}}}\left(\frac{x_{24}}{x_{23}}\right)^{\Delta_\epsilon}\times\\
&\Bigg({u_1'}^{\frac{\Delta _{\epsilon }}{2}}+{u_2'}^{\frac{\Delta _{\epsilon }}{2}}+\left(\frac{{u_1'} {u_2'}}{{v_1'} {v_2'}}\right)^{\Delta_{\sigma}} \left({v_1'}^{\frac{\Delta _{\epsilon }}{2}}+{v_2'}^{\frac{\Delta _{\epsilon }}{2}}\right)+\left(\frac{{u_1'} {u_2'}}{w'}\right)^{\Delta_{\sigma} } \left({w'}^{\frac{\Delta _{\epsilon }}{2}}+1\right)\Bigg)\,.
\end{split}
\end{equation}
Here, we treat $\Delta_{\sigma}$, $\Delta_{\epsilon}$, and $\lambda_{\sigma\sigma\epsilon}$ as free parameters and eventually input the data of the 3d critical Ising model. Upon expanding the given correlator in terms of the conformal blocks in the $(12)(45)$ channel, we find
\begin{equation}\label{mft-corr}
\begin{split}
\langle\sigma(x_1) \sigma(x_2) \epsilon(x_3) \sigma(x_4) \sigma(x_5) \rangle_{d} =\, &P(x_i)\Big(\lambda_{\sigma\sigma\epsilon}G^{(0)}_{(0,0,\Delta_\epsilon,0)} + \lambda_{\sigma\sigma\epsilon}G^{(0)}_{(\Delta_\epsilon,0,0,0)}\\
& + \sum_{n,l,n',l',n_{IJ}} \mathcal{P}(n,l,n',l',n_{IJ})G^{(n_{IJ})}_{(2\Delta_{\sigma}+2n+l,l,2\Delta_{\sigma}+2n'+l',l')}\Big)\,,
\end{split}
\end{equation} 
where $\mathcal{P}(n,l,n',l',n_{IJ}) = \lambda_{\sigma\sigma[\sigma,\sigma]_{n,l}}\lambda_{\sigma\sigma[\sigma,\sigma]_{n',l'}}\lambda^{n_{IJ}}_{[\sigma,\sigma]_{n,l}\epsilon[\sigma,\sigma]_{n',l'}}$ and the exchanged operators are interpreted as double-twist operators built out of $\sigma$, which can be schematically represented as $[\sigma,\sigma]_{n,l}\sim :\sigma \Box^{n}\partial_{\mu_{1}}\partial_{\mu_{2}}\ldots \partial_{\mu_{l}} \sigma:$. We note that only even-spin operators contribute to our correlator when expanding in the $(12)(45)$ channel.

We have computed a number of the OPE coefficients $P(n,l,n',l',n_{IJ})$ by expanding the given correlator in the OPE limits $x_{12}\to 0$ and $x_{45}\to 0$. Some of the coefficients are given by
\begin{align*}
\mathcal{P}(0,2,0,2,0)=\,&\frac{\Delta _{\sigma }^2 \Delta _{\epsilon }^2 \left(\Delta _{\epsilon }+2\right)^2 \lambda _{\sigma \sigma \epsilon }}{4 \left(2 \Delta _{\sigma }+1\right)^2}\,,\\
\mathcal{P}(0,2,0,4,0)=\,&\frac{\Delta _{\sigma }^2 \left(\Delta _{\sigma }+1\right) \Delta _{\epsilon }^2 \left(\Delta _{\epsilon }+2\right)^2 \left(\Delta _{\epsilon }+4\right) \left(\Delta _{\epsilon }+6\right) \lambda _{\sigma \sigma \epsilon }}{96 \left(2 \Delta _{\sigma }+1\right) \left(2 \Delta _{\sigma }+3\right) \left(2 \Delta _{\sigma }+5\right)}\,,\\
\mathcal{P}(0,4,0,4,4)=\,&\frac{\Delta _{\sigma } \left(\Delta _{\sigma }+1\right) \lambda _{\sigma \sigma \epsilon }}{96 \left(2 \Delta _{\sigma }+3\right)^2 \left(2 \Delta _{\sigma }+5\right)^2}(-4 \left(\Delta _{\sigma }+1\right) \left(\Delta _{\sigma } \left(4 \Delta _{\sigma }+15\right)+18\right) \Delta _{\epsilon }^3\\
&+64 \Delta _{\sigma } \left(\Delta _{\sigma }+1\right) \left(\Delta _{\sigma }+2\right) \left(\Delta _{\sigma }+3\right) \left(2 \Delta _{\sigma }+3\right) \left(2 \Delta _{\sigma }+5\right)+\Delta _{\sigma } \left(\Delta _{\sigma }+1\right) \Delta _{\epsilon }^4 \\
&-16 \left(\Delta _{\sigma } \left(\Delta _{\sigma } \left(2 \Delta _{\sigma } \left(8 \Delta _{\sigma }^2+74 \Delta _{\sigma }+263\right)+883\right)+681\right)+180\right) \Delta _{\epsilon }\\
& +4 \left(\Delta _{\sigma } \left(\Delta _{\sigma } \left(12 \Delta _{\sigma } \left(2 \Delta _{\sigma }+15\right)+515\right)+659\right)+324\right) \Delta _{\epsilon }^2)\,,\\
\end{align*}
\begin{align*}
\mathcal{P}(0,4,0,4,3)=\,&\frac{\Delta _{\sigma } \left(\Delta _{\sigma }+1\right) \Delta _{\epsilon }^2 \lambda _{\sigma \sigma \epsilon }}{24 \left(2 \Delta _{\sigma }+3\right)^2 \left(2 \Delta _{\sigma }+5\right)^2}(\Delta _{\sigma } \left(\Delta _{\sigma }+1\right) \Delta _{\epsilon }^3\\
&-6 \left(\Delta _{\sigma }+1\right) \left(\Delta _{\sigma } \left(2 \Delta _{\sigma }+7\right)+9\right) \Delta _{\epsilon }^2 \\
&+4 \left(\Delta _{\sigma } \left(2 \Delta _{\sigma } \left(6 \Delta _{\sigma } \left(\Delta _{\sigma }+7\right)+115\right)+281\right)+135\right) \Delta _{\epsilon }\\
&-16 \left(\Delta _{\sigma } \left(\Delta _{\sigma } \left(2 \Delta _{\sigma } \left(\Delta _{\sigma } \left(2 \Delta _{\sigma }+17\right)+55\right)+161\right)+96\right)+9\right))\,,\\
\mathcal{P}(0,4,0,4,2)=\,&\frac{\Delta _{\sigma } \left(\Delta _{\sigma }+1\right) \Delta_{\epsilon }^2 \left(\Delta _{\epsilon }+2\right)^2 \lambda _{\sigma \sigma \epsilon }}{32 \left(2 \Delta _{\sigma }+3\right)^2 \left(2 \Delta _{\sigma }+5\right)^2}(8 \left(\Delta _{\sigma } \left(\Delta _{\sigma }+3\right) \left(\Delta _{\sigma } \left(2 \Delta _{\sigma }+7\right)+13\right)+18\right)\\
&+\Delta _{\sigma } \left(\Delta _{\sigma }+1\right) \Delta _{\epsilon }^2-2 \left(\Delta _{\sigma }+1\right) \left(\Delta _{\sigma } \left(4 \Delta _{\sigma }+13\right)+18\right) \Delta _{\epsilon })\,,\\
\mathcal{P}(0,4,0,4,1)=\,&\frac{\Delta _{\sigma } \left(\Delta _{\sigma }+1\right)^2 \Delta _{\epsilon }^2 \left(\Delta _{\epsilon }+2\right)^2 \left(\Delta _{\epsilon }+4\right)^2 \left(\Delta _{\sigma } \left(\Delta _{\epsilon }-4 \left(\Delta _{\sigma }+3\right)\right)-18\right)\lambda _{\sigma \sigma \epsilon } }{144 \left(2 \Delta _{\sigma }+3\right)^2 \left(2 \Delta _{\sigma }+5\right)^2}\,,\nonumber\\
\mathcal{P}(0,4,0,4,0)=\,&\frac{\Delta _{\sigma }^2 \left(\Delta _{\sigma }+1\right)^2 \Delta _{\epsilon }^2 \left(\Delta _{\epsilon }+2\right)^2 \left(\Delta _{\epsilon }+4\right)^2 \left(\Delta _{\epsilon }+6\right)^2 \lambda _{\sigma \sigma \epsilon }}{2304 \left(2 \Delta _{\sigma }+3\right)^2 \left(2 \Delta _{\sigma }+5\right)^2}\,.
\end{align*}
All coefficients $\mathcal{P}(n,l,n',l',n_{IJ})$ with $2n+l+2n'+l'\leqslant 12$ are written in the attached {\fontfamily{lmss}\selectfont Mathematica} notebook.
Using the expression for the mean-field theory OPE coefficients $\lambda_{\sigma \sigma [\sigma,\sigma]_{n,l}}$ given by eq.~(\ref{mftopecoef}), together with the data from 3d critical Ising model (namely $\Delta_{\sigma}$, $\Delta_{\epsilon}$, and $\lambda_{\sigma \sigma \epsilon}$), we calculate $\lambda_{[\sigma,\sigma]_{0,l}\epsilon [\sigma,\sigma]_{0,l'}}^{n_{IJ}}$ for $l$ and $l'$ taking on the values $2,4,6,8$ (with $l+l'\leqslant 12$) and we write them in table~\ref{opetablemft}. We expect that these OPE coefficients roughly approximate the critical Ising OPE coefficients $\lambda_{\mathcal{O}_{\Delta,l}\epsilon \mathcal{O'}_{\Delta',l'}}^{n_{IJ}}$, where $\mathcal{O}_{\Delta,l}$ and $\mathcal{O'}_{\Delta',l'}$ are the operators of spins $l$ and $l'$ with the smallest critical dimension.

\begin{table}[h]
\begin{tabular}{|l|l|}
\hline
                                                                   & $\langle \sigma \sigma \epsilon \sigma \sigma \rangle_d$ \\ \hline
$\lambda_{[\sigma,\sigma]_{0,2}\epsilon[\sigma,\sigma]_{0,2}}^{2}$ & -0.353885      \\ \hline
$\lambda_{[\sigma,\sigma]_{0,2}\epsilon[\sigma,\sigma]_{0,2}}^{1}$ & -2.892385       \\ \hline
$\lambda_{[\sigma,\sigma]_{0,2}\epsilon[\sigma,\sigma]_{0,2}}^{0}$ & 0.988418       \\ \hline
$\lambda_{[\sigma,\sigma]_{0,2}\epsilon[\sigma,\sigma]_{0,4}}^{2}$ & 0.580279       \\ \hline
$\lambda_{[\sigma,\sigma]_{0,2}\epsilon[\sigma,\sigma]_{0,4}}^{1}$ & -3.100420       \\ \hline
$\lambda_{[\sigma,\sigma]_{0,2}\epsilon[\sigma,\sigma]_{0,4}}^{0}$ & 0.484382       \\ \hline
$\lambda_{[\sigma,\sigma]_{0,4}\epsilon[\sigma,\sigma]_{0,4}}^{4}$ & -0.439644      \\ \hline
$\lambda_{[\sigma,\sigma]_{0,4}\epsilon[\sigma,\sigma]_{0,4}}^{3}$ & -0.231194      \\ \hline
$\lambda_{[\sigma,\sigma]_{0,4}\epsilon[\sigma,\sigma]_{0,4}}^{2}$ & 3.849048        \\ \hline
$\lambda_{[\sigma,\sigma]_{0,4}\epsilon[\sigma,\sigma]_{0,4}}^{1}$ & -3.276278       \\ \hline
$\lambda_{[\sigma,\sigma]_{0,4}\epsilon[\sigma,\sigma]_{0,4}}^{0}$ & 0.237375       \\ \hline
\end{tabular}
\quad
\begin{tabular}{|l|l|}
\hline
                                                                   & $\langle \sigma \sigma \epsilon \sigma \sigma \rangle_d$ \\ \hline
$\lambda_{[\sigma,\sigma]_{0,2}\epsilon[\sigma,\sigma]_{0,6}}^{2}$ & 1.034441        \\ \hline
$\lambda_{[\sigma,\sigma]_{0,2}\epsilon[\sigma,\sigma]_{0,6}}^{1}$ & -2.351395       \\ \hline
$\lambda_{[\sigma,\sigma]_{0,2}\epsilon[\sigma,\sigma]_{0,6}}^{0}$ & 0.238792       \\ \hline
$\lambda_{[\sigma,\sigma]_{0,4}\epsilon[\sigma,\sigma]_{0,6}}^{4}$ & -0.426110      \\ \hline
$\lambda_{[\sigma,\sigma]_{0,4}\epsilon[\sigma,\sigma]_{0,6}}^{3}$ & -1.729458       \\ \hline
$\lambda_{[\sigma,\sigma]_{0,4}\epsilon[\sigma,\sigma]_{0,6}}^{2}$ & 5.210505        \\ \hline
$\lambda_{[\sigma,\sigma]_{0,4}\epsilon[\sigma,\sigma]_{0,6}}^{1}$ & -2.475238       \\ \hline
$\lambda_{[\sigma,\sigma]_{0,4}\epsilon[\sigma,\sigma]_{0,6}}^{0}$ & 0.117022       \\ \hline
$\lambda_{[\sigma,\sigma]_{0,6}\epsilon[\sigma,\sigma]_{0,6}}^{6}$ & -0.318259      \\ \hline
$\lambda_{[\sigma,\sigma]_{0,6}\epsilon[\sigma,\sigma]_{0,6}}^{5}$ & 0.419036       \\ \hline
$\lambda_{[\sigma,\sigma]_{0,6}\epsilon[\sigma,\sigma]_{0,6}}^{4}$ & 0.882017       \\ \hline
\end{tabular}
\quad
\begin{tabular}{|l|l|}
\hline
                                                                   & $\langle \sigma \sigma \epsilon \sigma \sigma \rangle_d$ \\ \hline
$\lambda_{[\sigma,\sigma]_{0,6}\epsilon[\sigma,\sigma]_{0,6}}^{3}$ & -5.958233       \\ \hline
$\lambda_{[\sigma,\sigma]_{0,6}\epsilon[\sigma,\sigma]_{0,6}}^{2}$ & 6.910388        \\ \hline
$\lambda_{[\sigma,\sigma]_{0,6}\epsilon[\sigma,\sigma]_{0,6}}^{1}$ & -1.868094       \\ \hline
$\lambda_{[\sigma,\sigma]_{0,6}\epsilon[\sigma,\sigma]_{0,6}}^{0}$ & 0.057690      \\ \hline
$\lambda_{[\sigma,\sigma]_{0,2}\epsilon[\sigma,\sigma]_{0,8}}^{2}$ & 1.076922        \\ \hline
$\lambda_{[\sigma,\sigma]_{0,2}\epsilon[\sigma,\sigma]_{0,8}}^{1}$ & -1.569468       \\ \hline
$\lambda_{[\sigma,\sigma]_{0,2}\epsilon[\sigma,\sigma]_{0,8}}^{0}$ & 0.118120       \\ \hline
$\lambda_{[\sigma,\sigma]_{0,4}\epsilon[\sigma,\sigma]_{0,8}}^{4}$ & -0.262017      \\ \hline
$\lambda_{[\sigma,\sigma]_{0,4}\epsilon[\sigma,\sigma]_{0,8}}^{3}$ & -2.891697       \\ \hline
$\lambda_{[\sigma,\sigma]_{0,4}\epsilon[\sigma,\sigma]_{0,8}}^{2}$ & 4.995170        \\ \hline
$\lambda_{[\sigma,\sigma]_{0,4}\epsilon[\sigma,\sigma]_{0,8}}^{1}$ & -1.649181       \\ \hline
$\lambda_{[\sigma,\sigma]_{0,4}\epsilon[\sigma,\sigma]_{0,8}}^{0}$ & 0.057886      \\ \hline
\end{tabular}
\caption{The OPE coefficients $\lambda_{[\sigma,\sigma]_{0,l}\epsilon[\sigma,\sigma]_{0,l'}}^{n_{IJ}}$ computed from the disconnected five-point correlator with $\Delta_{\sigma}$, $\Delta_{\epsilon}$, and $\lambda_{\sigma\sigma\epsilon}$ evaluated using the 3d critical Ising model data. These OPE coefficients roughly approximate the critical Ising OPE coefficients $\lambda_{\mathcal{O}_{\Delta,l}\epsilon \mathcal{O'}_{\Delta',l'}}^{n_{IJ}}$, where $\mathcal{O}_{\Delta,l}$ and $\mathcal{O'}_{\Delta',l'}$ are the operators of spins $l$ and $l'$ with the smallest conformal dimension.}\label{opetablemft}
\end{table}

With these results in place, it is straightforward to use the disconnected five-point correlator to approximate the truncated contributions in the numerical bootstrap of the five-point correlator in the 3d critical Ising model.

\section{Numerical bootstrap in the 3d critical Ising model}\label{five-point-numerics}

In this section we study the five-point correlator $\langle\sigma \sigma\epsilon\sigma\sigma \rangle$ in the 3d critical Ising model. We repeat the calculation performed in \cite{Poland:2023vpn}, but this time, we approximate the truncated primary operators by their counterparts from the disconnected five-point correlator. We observe  some nonneglible shifts in the computed values of OPE coefficients, indicating the importance of using an approximation scheme for the truncated part of the spectrum. 

Let us consider the aforementioned five-point correlator in the critical Ising model in the $(12)(45)$ OPE channel
\begin{equation}\label{ising-corr-split}
\begin{split}
\langle \sigma(x_1)\sigma(x_2)\epsilon(x_3)&\sigma(x_4)\sigma(x_5)\rangle =\lambda_{\sigma\sigma\epsilon}P(x_{i})G^{(0)}_{(0,0,\Delta_\epsilon,0)} + \lambda_{\sigma\sigma\epsilon}P(x_{i})G^{(0)}_{(\Delta_\epsilon,0,0,0)}  \\
&+\sum_{(\mathcal{O}_{\Delta,l},\mathcal{O'}_{\Delta',l'})\in \mathcal{S}}\sum_{n_{IJ}=0}^{{\rm min}(l,l')} P(x_i)\lambda_{\sigma \sigma \mathcal{O}_{\Delta,l}}\lambda_{\sigma \sigma \mathcal{O'}_{\Delta',l'}}\lambda_{\mathcal{O}_{\Delta,l}\epsilon \mathcal{O'}_{\Delta',l'}}^{n_{IJ}}G^{(n_{IJ})}_{(\Delta,l,\Delta',l')}\\
&+\sum_{(\mathcal{O}_{\Delta,l},\mathcal{O'}_{\Delta',l'})\notin \mathcal{S}}\sum_{n_{IJ}=0}^{{\rm min}(l,l')} P(x_i)\lambda_{\sigma \sigma \mathcal{O}_{\Delta,l}}\lambda_{\sigma \sigma \mathcal{O'}_{\Delta',l'}}\lambda_{\mathcal{O}_{\Delta,l}\epsilon \mathcal{O'}_{\Delta',l'}}^{n_{IJ}}G^{(n_{IJ})}_{(\Delta,l,\Delta',l')}\,,
\end{split}
\end{equation}
where we suppress the dependence of the conformal blocks on the cross-ratios. The set $\mathcal{S}$ contains the pairs of operators we explicitly include in the bootstrap; the contributions of the rest we approximate by the disconnected  five-point correlator. Initially, we will take the set $\mathcal{S}$ to contain all pairs of operators $\{\epsilon, \epsilon', T_{\mu\nu}, C_{\mu\nu\rho\sigma}\}$, so that  $\Delta_{\rm cutoff}= \Delta_{C}\approx 5.022$. Later, we will be adding more contributions to $\mathcal{S}$  in order to test the sensitivity of the obtained OPE coefficients to our approximation with the disconnected correlator. 

Now, the crossing relation can be written as
\begin{equation}
\begin{split}
&\sum_{(\mathcal{O}_{\Delta,l},\mathcal{O'}_{\Delta',l'})\in \mathcal{S}}\sum_{n_{IJ}=0}^{{\rm min}(l,l')} \frac{\lambda_{\sigma\sigma \mathcal{O}_{\Delta,l}} \lambda_{\sigma\sigma \mathcal{O'}_{\Delta',l'}}\lambda_{\mathcal{O}_{\Delta,l}\epsilon \mathcal{O'}_{\Delta',l'}}^{n_{IJ}}}{\lambda_{\sigma\sigma\epsilon}}\mathcal{F}^{n_{IJ}}_{\Delta,l,\Delta',l'}\\
&+\sum_{(\mathcal{O}_{\Delta,l},\mathcal{O'}_{\Delta',l'})\notin \mathcal{S}}\sum_{n_{IJ}=0}^{{\rm min}(l,l')} \frac{\lambda_{\sigma\sigma \mathcal{O}_{\Delta,l}} \lambda_{\sigma\sigma \mathcal{O'}_{\Delta',l'}}\lambda_{\mathcal{O}_{\Delta,l}\epsilon \mathcal{O'}_{\Delta',l'}}^{n_{IJ}}}{\lambda_{\sigma\sigma\epsilon}}\mathcal{F}^{n_{IJ}}_{\Delta,l,\Delta',l'}-1=0\,,
\end{split}
\end{equation}
where 
\begin{equation}
\begin{split}
&\mathcal{F}^{n_{IJ}}_{\Delta, l, \Delta', l'}\equiv\\
&\frac{(v_1' v_2')^{\Delta_{\sigma}}G^{(n_{IJ})}_{(\Delta, l, \Delta', l')}(u_1',v_1',u_2',v_2',w')-(u_1' u_2')^{\Delta_{\sigma}}G^{(n_{IJ})}_{(\Delta, l, \Delta', l')}(v_1',u_1',v_2',u_2',w')}{(u_1' u_2')^{\Delta_{\sigma}}({v_1'}^{\Delta_{\epsilon}/2}+{v_2'}^{\Delta_{\epsilon}/2})-(v_1' v_2')^{\Delta_{\sigma}}({u_1'}^{\Delta_{\epsilon}/2}+{u_2'}^{\Delta_{\epsilon}/2})}\,.
\end{split}
\end{equation}
Analogously, we write the disconnected correlator~(\ref{mft-corr}) as
\begin{equation}\label{mft-corr-split}
\begin{split}
&\langle \sigma(x_1)\sigma(x_2)\epsilon(x_3)\sigma(x_4)\sigma(x_5)\rangle_{d} =\lambda_{\sigma\sigma \epsilon}^{d}P(x_i)G^{(0)}_{(0,0,\Delta_\epsilon^{d},0)} + \lambda_{\sigma\sigma \epsilon}^{d}P(x_i)G^{(0)}_{(\Delta_\epsilon^{d},0,0,0)}\\
&+ \sum_{([\sigma,\sigma]_{n,l},[\sigma,\sigma]_{n',l'})\in \mathcal{S}_{d}}\sum_{n_{IJ}=0}^{{\rm min}(l,l')} P(x_i)\mathcal{P}(n,l,n',l',n_{IJ})G^{(n_{IJ})}_{(2\Delta_\sigma^{d}+2n+l,l,2\Delta_\sigma^{d}+2n'+l',l')}\\
&+ \sum_{([\sigma,\sigma]_{n,l},[\sigma,\sigma]_{n',l'})\notin \mathcal{S}_{d}}\sum_{n_{IJ}=0}^{{\rm min}(l,l')} P(x_i)\mathcal{P}(n,l,n',l',n_{IJ})G^{(n_{IJ})}_{(2\Delta_\sigma^{d}+2n+l,l,2\Delta_\sigma^{d}+2n'+l',l')}\,,
\end{split}
\end{equation}
where we denote the quantum numbers in the disconnected correlator with superscript $d$.  When the set $\mathcal{S}$ in the critical Ising model contains all pairs of operators $\{\epsilon, \epsilon', T_{\mu\nu}, C_{\mu\nu\rho\sigma}\}$, the set $\mathcal{S}_{d}$ contains all pairs of operators $\{[\sigma,\sigma]_{0,0},[\sigma,\sigma]_{1,0},[\sigma,\sigma]_{0,2}, [\sigma,\sigma]_{0,4}\}$. Note that the identity contribution is explicitly included both in eq.~(\ref{ising-corr-split}) and eq.~(\ref{mft-corr-split}). 

Now, the approximation we make is
\begin{equation}\label{mft-approx}
\begin{split}
&\sum_{(\mathcal{O}_{\Delta,l},\mathcal{O'}_{\Delta',l'})\notin \mathcal{S}}\sum_{n_{IJ}=0}^{{\rm min}(l,l')} \frac{\lambda_{\sigma\sigma \mathcal{O}_{\Delta,l}} \lambda_{\sigma\sigma \mathcal{O'}_{\Delta',l'}}\lambda_{\mathcal{O}_{\Delta,l}\epsilon \mathcal{O'}_{\Delta',l'}}^{n_{IJ}}}{\lambda_{\sigma\sigma\epsilon}}\mathcal{F}^{n_{IJ}}_{\Delta,l,\Delta',l'}\approx\\
&\sum_{([\sigma,\sigma]_{n,l},[\sigma,\sigma]_{n',l'})\notin \mathcal{S}_{d}}\sum_{n_{IJ}=0}^{{\rm min}(l,l')} \frac{\mathcal{P}(n,l,n',l',n_{IJ})}{\lambda_{\sigma\sigma\epsilon}^{d}}\tilde{\mathcal{F}}^{n_{IJ}}_{2\Delta_\sigma^{d}+2n+l,l,2\Delta_\sigma^{d}+2n'+l',l'}\,,
\end{split}
\end{equation}
where 
\begin{equation}\label{Fdefmft}
\begin{split}
\tilde{\mathcal{F}}^{n_{IJ}}_{2\Delta_{\sigma}^{d}+2n+l, l, 2\Delta_{\sigma}^{d}+2n'+l', l'}\equiv\,
&\frac{(v_1' v_2')^{\Delta_{\sigma}^{d}}G^{(n_{IJ})}_{(2\Delta_{\sigma}^{d}+2n+l, l, 2\Delta_{\sigma}^{d}+2n'+l', l')}(u_1',v_1',u_2',v_2',w')}{(u_1' u_2')^{\Delta_{\sigma}^{d}}({v_1'}^{\Delta_{\epsilon}^{d}/2}+{v_2'}^{\Delta_{\epsilon}^{d}/2})-(v_1' v_2')^{\Delta_{\sigma}^{d}}({u_1'}^{\Delta_{\epsilon}^{d}/2}+{u_2'}^{\Delta_{\epsilon}^{d}/2})}\\
&-\frac{(u_1' u_2')^{\Delta_{\sigma}^{MFT}}G^{(n_{IJ})}_{(2\Delta_{\sigma}^{d}+2n+l, l, 2\Delta_{\sigma}^{d}+2n'+l', l')}(v_1',u_1',v_2',u_2',w')}{(u_1' u_2')^{\Delta_{\sigma}^{d}}({v_1'}^{\Delta_{\epsilon}^{d}/2}+{v_2'}^{\Delta_{\epsilon}^{d}/2})-(v_1' v_2')^{\Delta_{\sigma}^{d}}({u_1'}^{\Delta_{\epsilon}^{d}/2}+{u_2'}^{\Delta_{\epsilon}^{d}/2})}\,.
\end{split}
\end{equation}
The approximate crossing relation can then be written as
\begin{equation}\label{cross-rel}
\begin{split}
&\sum_{(\mathcal{O}_{\Delta,l},\mathcal{O'}_{\Delta',l'})\in \mathcal{S}}\sum_{n_{IJ}=0}^{{\rm min}(l,l')} \frac{\lambda_{\sigma\sigma \mathcal{O}_{\Delta,l}} \lambda_{\sigma\sigma \mathcal{O'}_{\Delta',l'}}\lambda_{\mathcal{O}_{\Delta,l}\epsilon \mathcal{O'}_{\Delta',l'}}^{n_{IJ}}}{\lambda_{\sigma\sigma\epsilon}}\mathcal{F}^{n_{IJ}}_{\Delta,l,\Delta',l'}\\
&-\sum_{([\sigma,\sigma]_{n,l},[\sigma,\sigma]_{n',l'})\in \mathcal{S}_{d}}\sum_{n_{IJ}=0}^{{\rm min}(l,l')} \frac{\mathcal{P}(n,l,n',l',n_{IJ})}{\lambda_{\sigma\sigma\epsilon}^{d}}\tilde{\mathcal{F}}^{n_{IJ}}_{2\Delta_\sigma^{d}+2n+l,l,2\Delta_\sigma^{d}+2n'+l',l'}=0\,.
\end{split}
\end{equation}
We fix the following data in the critical Ising model \cite{Simmons-Duffin:2016wlq, Su:unpublised}
\begin{equation}\label{ising-data}
\begin{split}
&\Delta_\epsilon = 1.412625\,, \quad \Delta_{\epsilon'} = 3.82968\,, \quad \Delta_C = 5.022665\,, \quad \lambda_{\sigma\sigma\epsilon}=1.0518537\,,\\
& \lambda_{\epsilon\epsilon\epsilon}=1.532435\,, \quad \lambda_{\sigma\sigma\epsilon'}=0.053012\,,\quad \lambda_{\epsilon\epsilon\epsilon'}=1.5360\,,\quad \lambda_{\epsilon' \epsilon \epsilon'}=2.3955808\,\\
&\lambda_{\sigma\sigma T}=0.65227552\,,\quad \lambda_{\epsilon\epsilon T}= 1.7782942\,, \quad \lambda_{\sigma\sigma C}=0.276304\,, \quad \lambda_{\epsilon\epsilon C}=0.99168\,.
\end{split}
\end{equation}
The unknown OPE coefficients are $(\lambda_{T\epsilon T}^{0}$, $\lambda_{T\epsilon C}^{0}$, $\lambda_{C\epsilon C}^{n_{IJ}}$, $\lambda_{\epsilon' \epsilon C})$. Additionally, in the critical Ising model, we treat $\Delta_{\sigma}$ as an unknown parameter as doing so allows for a convenient method for choosing the derivative sets. In the part of eq.~(\ref{cross-rel}) that comes from the disconnected correlator, we do not leave any free parameters, namely, we fix both $\Delta_{\sigma}^{d}=0.5181489$ and $\Delta_{\epsilon}^{d}=\Delta_{\epsilon}$.\footnote{Note that $\lambda_{\sigma\sigma\epsilon}^{d}$ doesn't enter the mean-field theory calculation since $\mathcal{P}(n,l,n',l',n_{IJ})\propto \lambda_{\sigma\sigma\epsilon}^{d}$.}

Now, we take derivatives of eq.~(\ref{cross-rel}) and evaluate them at the configuration given by
\begin{equation}\label{configuration}
u_1'=v_1'=u_2'=v_2'=1\,, \qquad w'=\frac{3}{2}\,,
\end{equation}
or in radial coordinates
\begin{equation}
R=2-\sqrt{3}\,, \qquad r=1\,, \qquad \eta_1=\eta_2=\hat{w}=0\,,
\end{equation}
to obtain linearly independent constraints for the unknown OPE coefficients in the 3d critical Ising model.

The equations obtained by taking derivatives of eq.~(\ref{cross-rel}) are denoted by $e_i(\Delta_{1},\lambda)$, with
\begin{equation}\label{equations}
\begin{split}
 e_{i}(\Delta_{\sigma},\lambda)=\,&
 \mathcal{D}_{i}\Bigg(\sum_{(\mathcal{O}_{\Delta,l},\mathcal{O'}_{\Delta',l'})\in \mathcal{S}}\sum_{n_{IJ}=0}^{{\rm min}(l,l')} \frac{\lambda_{\sigma\sigma \mathcal{O}_{\Delta,l}} \lambda_{\sigma\sigma \mathcal{O'}_{\Delta',l'}}\lambda_{\mathcal{O}_{\Delta,l}\epsilon \mathcal{O'}_{\Delta',l'}}^{n_{IJ}}}{\lambda_{\sigma\sigma\epsilon}}\mathcal{F}^{n_{IJ}}_{\Delta,l,\Delta',l'}\\
&-\sum_{([\sigma,\sigma]_{n,l},[\sigma,\sigma]_{n',l'})\in \mathcal{S}_{d}}\sum_{n_{IJ}=0}^{{\rm min}(l,l')} \frac{\mathcal{P}(n,l,n',l',n_{IJ})}{\lambda_{\sigma\sigma\epsilon}^{d}}\tilde{\mathcal{F}}^{n_{IJ}}_{2\Delta_\sigma^{d}+2n+l,l,2\Delta_\sigma^{d}+2n'+l',l'}   \Bigg)\Bigg|_{(\ref{configuration})}\,,
\end{split}
\end{equation}
where $i\geqslant1$ and $\mathcal{D}_{i}$ represent derivatives with respect to the cross-ratio parametrization given in appendix~\ref{parametrization}. We consider all constraints obtained by taking up to three derivatives. The label $e_{1}$ represents the equation obtained by not taking any derivatives of eq.~(\ref{cross-rel}), $\mathcal{D}_{1}\equiv 1$. Lastly, $\lambda$ represents all unknown OPE coefficients. 

We define the cost functions
\begin{equation}
f_{\{r_i\}}(\Delta_{\sigma},\lambda)= \sum_{i=1}^{\mathcal{C}}r_i\left(\frac{e_{i}(\Delta_{\sigma},\lambda)}{e_{i}(\Delta_{\sigma},0)}\right)^2\,,
\end{equation}
where $r_{i}$ are (pseudo-)randomly generated real numbers, $r_{i}\in [0,1]$, which give random weights to each term.
The $i$ sum runs over the set of derivatives $\mathcal{D}=\{\mathcal{D}_{i}|i=1,2,\ldots, \mathcal{C}\}$ used to produce the constraints $e_{i}(\Delta_{\sigma},\lambda)$. Now, for each set of numbers $r_i$, we minimize the function $f_{\{r_i\}}(\Delta_{\sigma},\lambda)$ to compute $\Delta_{\sigma}$ and the unknown OPE coefficients $\lambda$. We subsequently apply the same procedure for different values of $\mathcal{C}$. For each $\mathcal{C}$, we select the set of constraints $e_i(\Delta_\sigma,\lambda)$ that gives $\Delta_{\sigma}$ closest to the known value with the smallest standard deviation. We ultimately merge the data obtained with different values of $\mathcal{C}$ into a single set. We then take the average of the resulting values of $\Delta_\sigma$ and the unknown OPE coefficients, using their standard deviations as a rough estimate of our error bars. 

\subsection{Numerical results}

Let us first discuss the case when the set $\mathcal{S}$ is defined as 
\begin{equation}\label{sdef}
\mathcal{S}\equiv\{(\mathcal{O}_{\Delta,l},\mathcal{O'}_{\Delta',l'})|\mathcal{O}_{\Delta,l}, \mathcal{O'}_{\Delta',l'}\in\{\epsilon, \epsilon', T_{\mu\nu},C_{\mu\nu\rho\sigma}\}\}\,.
\end{equation}
Then, we define the set $\mathcal{S}_{MFT}$ as
\begin{equation}\label{smftdef}
\mathcal{S}_{MFT}\equiv\{(\mathcal{O}_{\Delta,l},\mathcal{O'}_{\Delta',l'})|\mathcal{O}_{\Delta,l}, \mathcal{O'}_{\Delta',l'}\in\{[\sigma,\sigma]_{0,0}, [\sigma,\sigma]_{1,0}, [\sigma,\sigma]_{0,2}, [\sigma,\sigma]_{0,4}\}\}\,.
\end{equation}

\begin{figure}[H]
\centering
\includegraphics[width=0.315\textwidth]{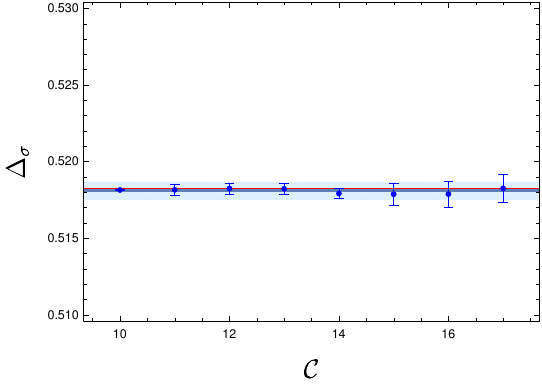}
\hspace{0.01\textwidth}
\includegraphics[width=0.315\textwidth]{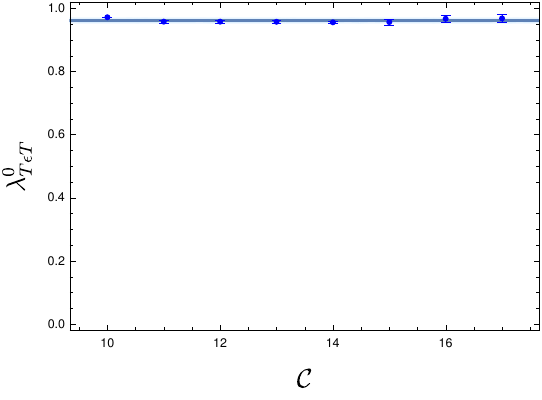}
\hspace{0.01\textwidth}
\includegraphics[width=0.315\textwidth]{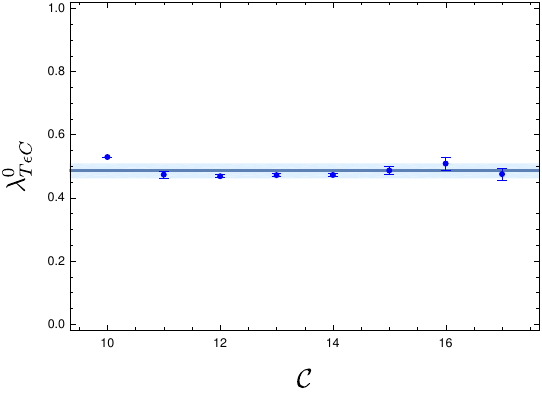}\\
\includegraphics[width=0.315\textwidth]{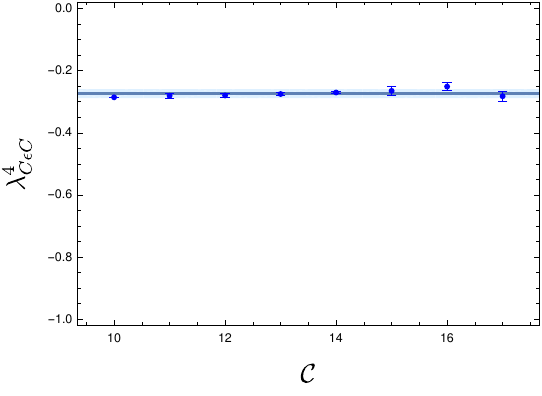}
\hspace{0.01\textwidth}
\includegraphics[width=0.315\textwidth]{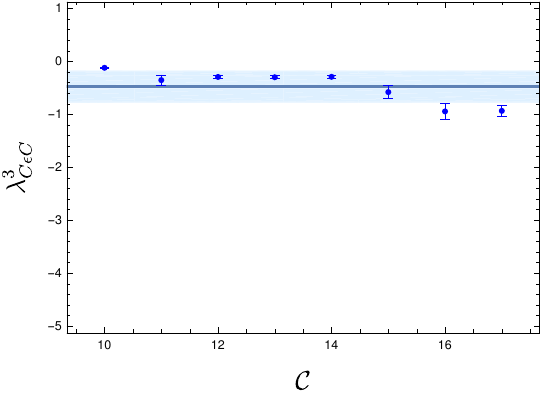}
\hspace{0.01\textwidth}
\includegraphics[width=0.315\textwidth]{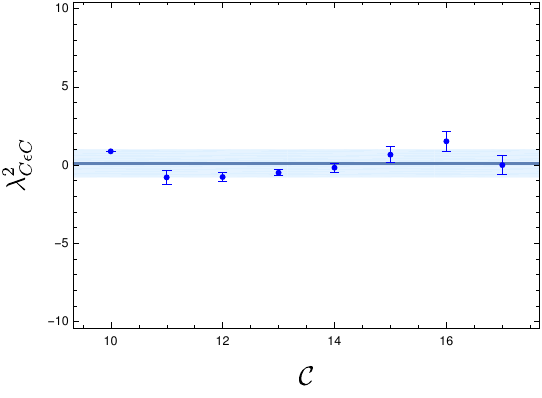}\\
\includegraphics[width=0.315\textwidth]{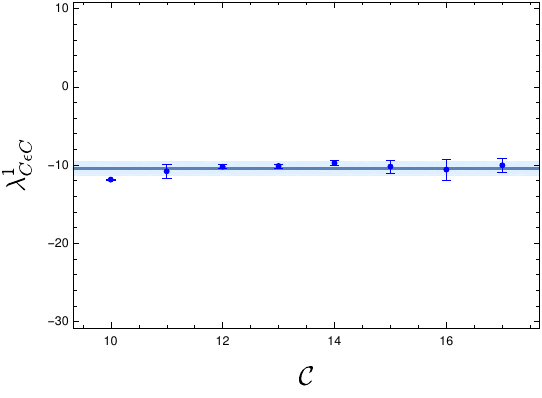}
\hspace{0.01\textwidth}
\includegraphics[width=0.315\textwidth]{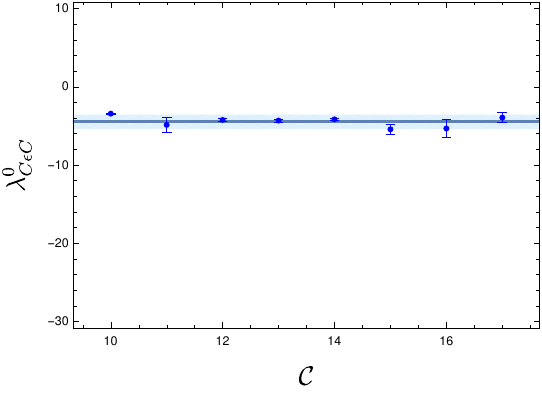}
\hspace{0.01\textwidth}
\includegraphics[width=0.315\textwidth]{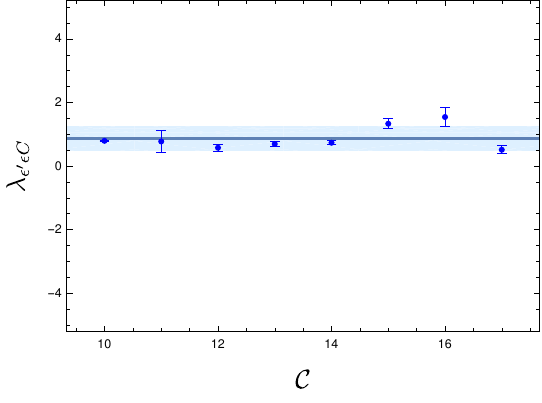}
\caption{{\small $\Delta_{\sigma}$ and the unknown OPE coefficients in the critical 3d Ising CFT, computed by minimizing the cost functions $f_{\{r_i\}}(\Delta_{\phi},\lambda)$, defined with different choices of the number of constraints $\mathcal{C}$. The sets $\mathcal{S}$ and $\mathcal{S}_{MFT}$ are as given by eq.~(\ref{sdef}) and eq.~(\ref{smftdef}). For each value of $\mathcal{C}$, the set of derivatives that we use is chosen such that $\Delta_{\sigma}$ is closest to the known value with the minimal standard deviation after averaging over data obtained by randomly selecting the weights $r_i$ in the cost function. The sets are given explicitly in eq.~(\ref{setsv1}). The horizontal red line represents the best known value of $\Delta_{\sigma}$. The horizontal blue lines are the mean values of the data and the blue strips represent the error bars given in table \ref{numrestab}.   One can observe that the standard deviations increase as we increase $\mathcal{C}$. The scales of the vertical axes are the same as those in \cite{Poland:2023vpn}.}}
\label{isingv1-plots}
\end{figure}

We have 8 unknown OPE coefficients and we also treat $\Delta_{\sigma}$ in the critical Ising model as a free parameter, so we use $\mathcal{C}\geqslant10$ constraints in the cost function. The choice of derivatives we use to generate the constraints that give $\Delta_{\sigma}$ closest to its known value is given by eq.~(\ref{setsv1}).

Next, we will add the contribution $(\epsilon, S_{\mu\nu\rho\sigma\alpha\beta})$ to $\mathcal{S}$, and the contribution $([\sigma,\sigma]_{0,0}, [\sigma,\sigma]_{0,6})$ to $\mathcal{S}_{d}$. The OPE data for the spin-6 contribution in the critical Ising model are given by \cite{Simmons-Duffin:2016wlq}
\begin{equation}
\Delta_{S}=7.028488\,, \quad \lambda_{\sigma\sigma S}=0.1259328\,, \quad \lambda_{\epsilon\epsilon S}=0.529088\,.
\end{equation}

\begin{figure}[H]
\centering
\includegraphics[width=0.315\textwidth]{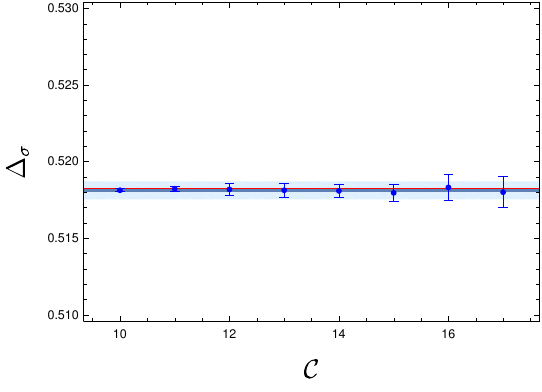}
\hspace{0.01\textwidth}
\includegraphics[width=0.315\textwidth]{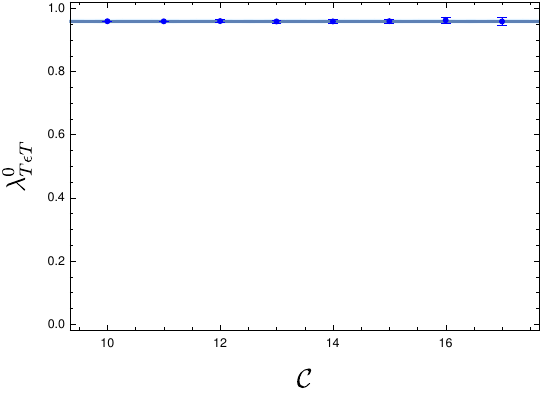}
\hspace{0.01\textwidth}
\includegraphics[width=0.315\textwidth]{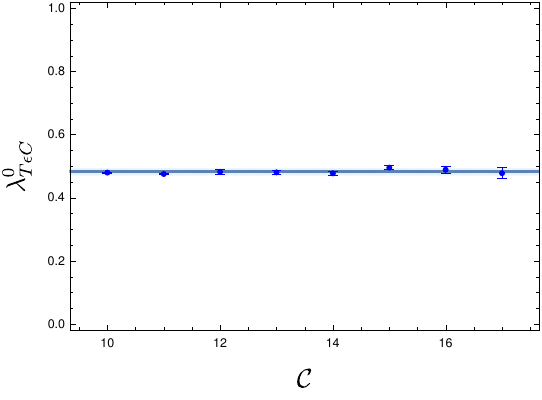}\\
\includegraphics[width=0.315\textwidth]{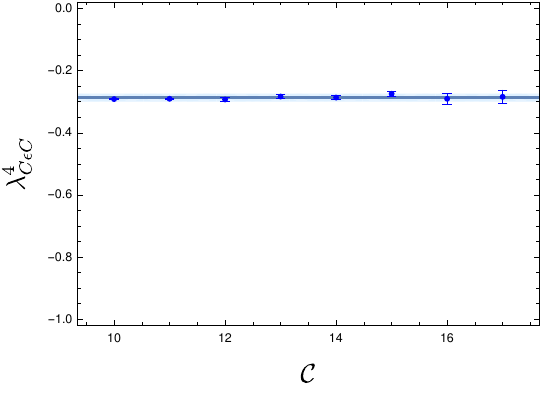}
\hspace{0.01\textwidth}
\includegraphics[width=0.315\textwidth]{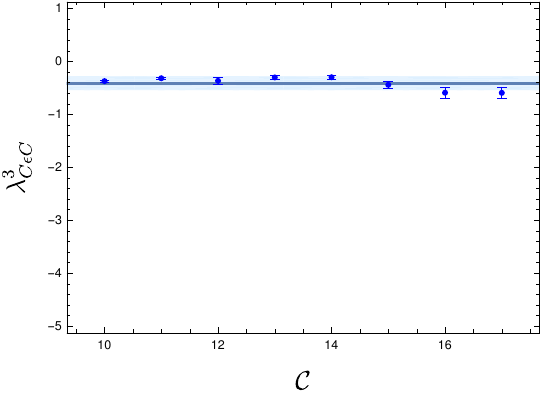}
\hspace{0.01\textwidth}
\includegraphics[width=0.315\textwidth]{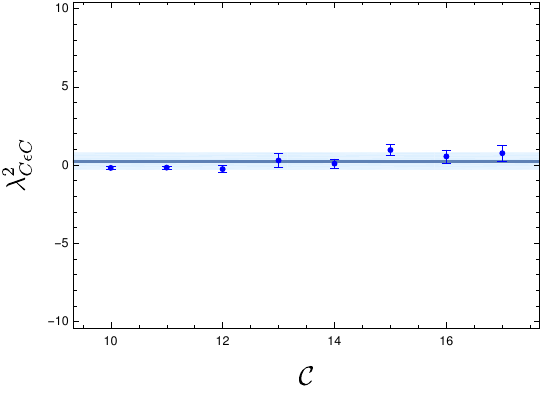}\\
\includegraphics[width=0.315\textwidth]{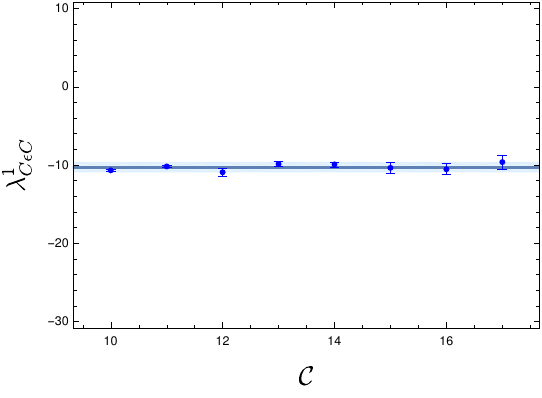}
\hspace{0.01\textwidth}
\includegraphics[width=0.315\textwidth]{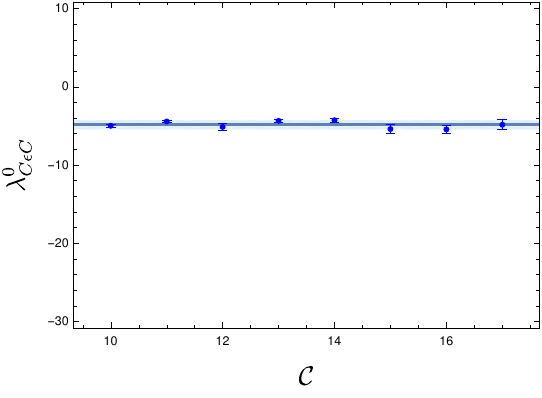}
\hspace{0.01\textwidth}
\includegraphics[width=0.315\textwidth]{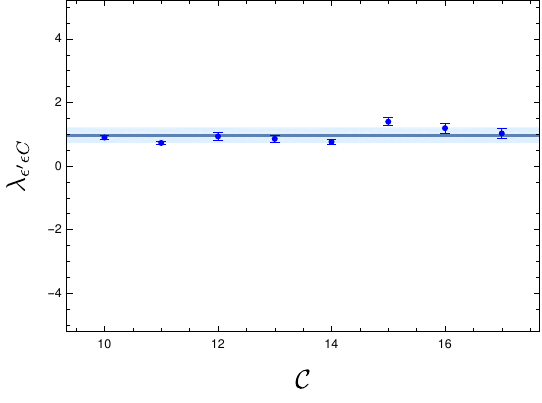}
\caption{{\small $\Delta_{\sigma}$ and the unknown OPE coefficients in the critical 3d Ising CFT, computed by minimizing the cost functions $f_{\{r_i\}}(\Delta_{\phi},\lambda)$, defined with different choices of the number of constraints $\mathcal{C}$. The sets $\mathcal{S}$ and $\mathcal{S}_{MFT}$ are as given by eq.~(\ref{sdef}) and eq.~(\ref{smftdef}), with the addition of $(\epsilon, S_{\mu\nu\rho\sigma\alpha\beta})$ and $([\sigma,\sigma]_{0,0},[\sigma,\sigma]_{0,6})$, respectively. For each value of $\mathcal{C}$, the set of derivatives that we use is chosen such that $\Delta_{\sigma}$ is closest to the known value with the minimal standard deviation after averaging over data obtained by randomly selecting the weights $r_i$ in the cost function. The sets are given explicitly in eq.~(\ref{setsv2}).  The horizontal red line represents the best known value of $\Delta_{\sigma}$. The horizontal blue lines are the mean values of the data and the blue strips represent the error bars given in table \ref{numrestab}.   One can observe that the standard deviations increase as we increase $\mathcal{C}$. The scales of the vertical axes are the same as those in \cite{Poland:2023vpn}.}}
\label{isingv2-plots}
\end{figure}
\noindent Adding this contribution does not increase the number of unknown OPE coefficients in the problem. A reason for adding this contribution (and others like it with higher spin) is the fact that the double-trace contribution in the disconnected correlator $[\sigma,\sigma]_{0,0}$ does not approximate well the $\epsilon$-contributions in the critical Ising model, due to its large anomalous dimension. 

We can further add $(\epsilon, \mathcal{E}_{\mu_{1}\ldots \mu_{8}})$ in $\mathcal{S}$ and $([\sigma,\sigma]_{0,0},[\sigma,\sigma]_{0,8})$ in $\mathcal{S}_{MFT}$. The OPE data of the spin-8 operator is given by \cite{Simmons-Duffin:2016wlq}
\begin{equation}
\Delta_{\mathcal{E}}=9.031023\,, \quad \lambda_{\sigma\sigma\mathcal{E}}=0.05896\,, \quad \lambda_{\epsilon\epsilon\mathcal{E}}=0.277088\,.
\end{equation}

\begin{figure}[H]
\centering
\includegraphics[width=0.315\textwidth]{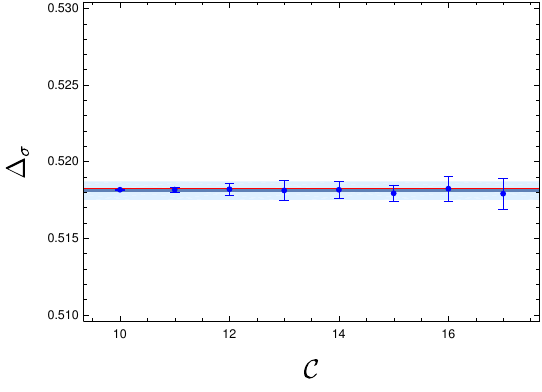}
\hspace{0.01\textwidth}
\includegraphics[width=0.315\textwidth]{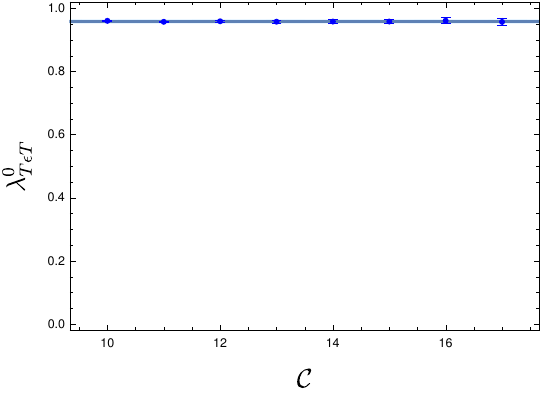}
\hspace{0.01\textwidth}
\includegraphics[width=0.315\textwidth]{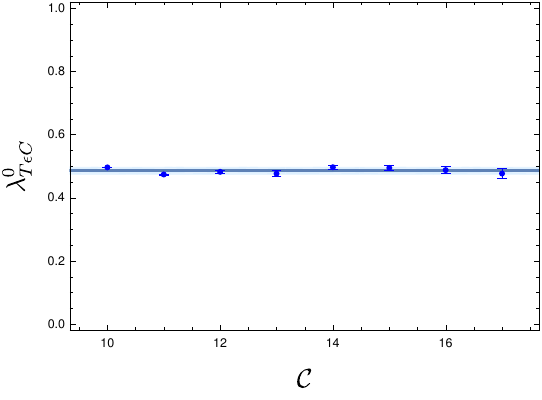}\\
\includegraphics[width=0.315\textwidth]{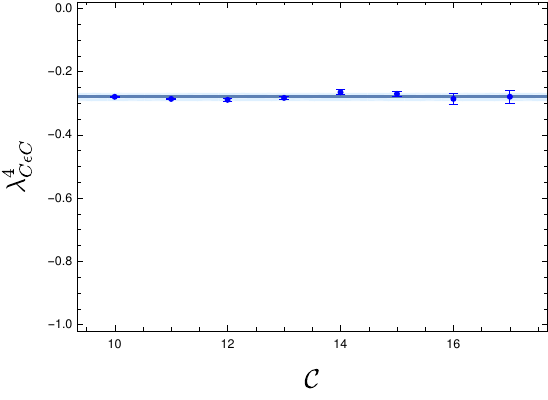}
\hspace{0.01\textwidth}
\includegraphics[width=0.315\textwidth]{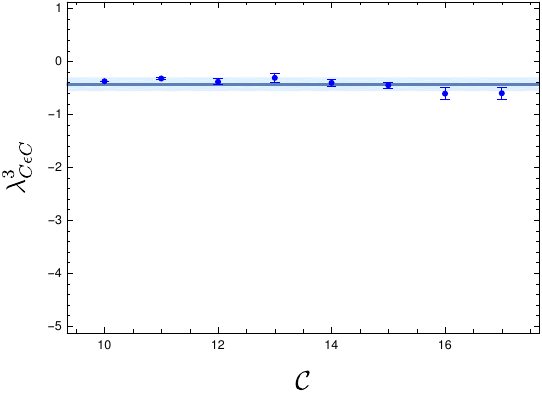}
\hspace{0.01\textwidth}
\includegraphics[width=0.315\textwidth]{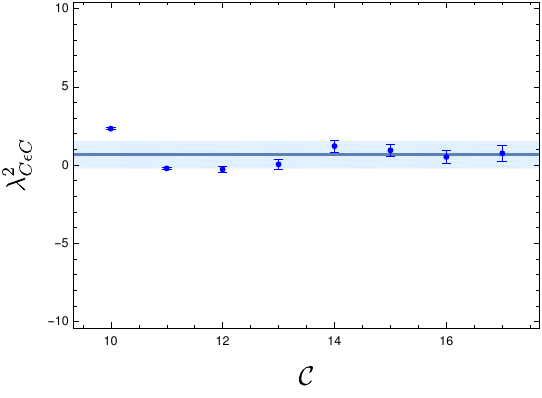}\\
\includegraphics[width=0.315\textwidth]{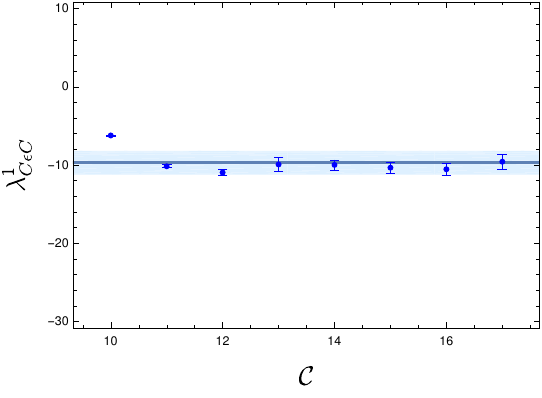}
\hspace{0.01\textwidth}
\includegraphics[width=0.315\textwidth]{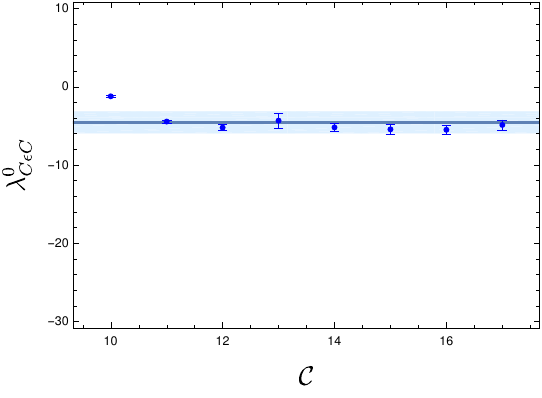}
\hspace{0.01\textwidth}
\includegraphics[width=0.315\textwidth]{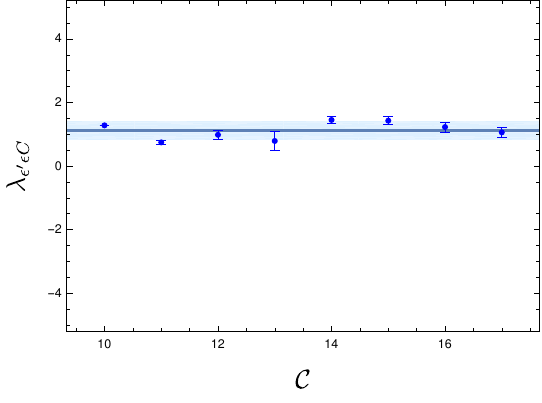}
\caption{{\small $\Delta_{\sigma}$ and the unknown OPE coefficients in the critical 3d Ising CFT computed by minimizing the cost functions $f_{\{r_i\}}(\Delta_{\phi},\lambda)$ defined with different choices of the number of constraints $\mathcal{C}$. The sets $\mathcal{S}$ and $\mathcal{S}_{MFT}$ are as given by eq.~(\ref{sdef}) and eq.~(\ref{smftdef}), with the addition of $(\epsilon, S_{\mu\nu\rho\sigma\alpha\beta})$ and $(\epsilon, \mathcal{E}_{\mu_1\ldots \mu_8})$ in $\mathcal{S}$ and $([\sigma,\sigma]_{0,0},[\sigma,\sigma]_{0,6})$ and $([\sigma,\sigma]_{0,0},[\sigma,\sigma]_{0,8})$ in $\mathcal{S}_{MFT}$. For each value of $\mathcal{C}$, the set of derivatives that we use is chosen such that $\Delta_{\sigma}$ is closest to the known value with the minimal standard deviation after averaging over data obtained by randomly selecting the weights $r_i$ in the cost function. The sets are explicitly given by eq.~(\ref{setsv3}). The horizontal red line represents the best known value of $\Delta_{\sigma}$. The horizontal blue lines are the mean values of the data and the blue strips represent the error bars given in table \ref{numrestab}.   One can observe that the standard deviations increase as we increase $\mathcal{C}$. The scales of the vertical axes are the same as those in \cite{Poland:2023vpn}.}}
\label{isingv3-plots}
\end{figure}

Again, this contribution does not introduce any new unknown OPE coefficients into the system. In principle, one can add an arbitrary number of contributions $(\epsilon, \mathcal{O'}_{\Delta',l'})$, but the results of the bootstrap algorithm quickly become insensitive to additional contributions, as can be inferred already when we add the spin-8 operator together with $\epsilon$.

\begin{table}[]
\centering
\begin{tabular}{|l|l|l|l|l|}
\hline
                                 & ${\rm no\, disc.}$ \cite{Poland:2023vpn} & ${\rm  \mathcal{S}\,in}$ eq.~(\ref{sdef}) & $+(\epsilon, S_{\mu\nu\rho\sigma\alpha\delta})$ & $+(\epsilon, \mathcal{E}_{\mu_{1}\ldots \mu_{8}})$ \\ \hline
$\Delta_{\sigma}$                & 0.518(2)         & 0.5181(6)                                     & 0.5181(6)                                       & 0.5181(7)                                                   \\ \hline
$\lambda_{T\epsilon T}^{0}$      & 0.81(5)          & 0.96(1)                                       & 0.959(8)                                        & 0.958(7)                                                   \\
$\lambda_{T\epsilon C}^{0}$      & 0.30(6)          & 0.48(3)                                       & 0.48(2)                                         &  0.48(2)                                                  \\
$\lambda_{C\epsilon C}^{4}$      & -0.3(1)          & -0.27(2)                                      & -0.28(2)                                        & -0.28(2)                                                   \\
$\lambda_{C\epsilon C}^{3}$      & -2(2)            & -0.5(4)                                       & -0.4(2)                                         &  -0.4(2)                                                  \\
$\lambda_{C\epsilon C}^{2}$      & 2(5)             & 0.1(9)                                        & 0.3(6)                                          &   0.6(9)                                                 \\
$\lambda_{C\epsilon C}^{1}$      & -5(11)           & -10(1)                                        & -10.3(7)                                        &      -10(2)                                              \\
$\lambda_{C\epsilon C}^{0}$      & -3(11)           & -4(1)                                         & -4.9(6)                                         & -4(2)                                                    \\ 
$\lambda_{\epsilon' \epsilon C}$ & 0(2)             & 0.9(4)                                        & 1.0(3)                                          &     1.1(3)                                               \\ \hline
\end{tabular}
\caption{Numerical data for $\Delta_{\sigma}$ and unknown OPE coefficients obtained in \cite{Poland:2023vpn} without or with the disconnected correlator approximation, for different choices of the sets $\mathcal{S}$ and $\mathcal{S}_{MFT}$.}\label{numrestab}
\end{table}

Our results are summarized in table~\ref{numrestab}. We remark that they are broadly consistent with the previous computations of~\cite{Poland:2023vpn}, but we observe a noticeable upward shift in the central values of the $\lambda_{T\epsilon T}^0$ and $\lambda_{T \epsilon C}^0$ coefficients. In addition, we find that our results for both these and the $\lambda_{C\epsilon C}^{n_{IJ}}$ coefficients have smaller errors. We note that our new value of the $\lambda_{T\epsilon T}^{0}$ OPE coefficient still satisfies the bounds set in \cite{Cordova:2017zej} ($|\lambda_{T\epsilon T}^0| \simeq 0.958(7) \leq 0.981(2)$), and is close to but somewhat larger than the value $\lambda_{T\epsilon T}^0 \simeq 0.9162(73)$ computed in~\cite{Hu:2023xak} using the fuzzy sphere regularization approach~\cite{Zhu:2022gjc, Hu:2023xak, Lao:2023zis, Han:2023yyb}. It will be interesting to study how this OPE coefficient is affected by subleading $1/N$ corrections in the fuzzy sphere method. It will also be enlightening to test the consistency of these determinations against future rigorous results from the 4-point bootstrap for spinning operators.

\section{Discussion}\label{disc-sec}

In this paper we presented a new algorithm for the numerical evaluation of five-point conformal blocks, similar to the method used for four-point blocks in \cite{Costa:2016xah}; namely, we write a set of recursion relations for the coefficients in the ansatz for the five-point blocks (as a series expansion in radial coordinates) and then solve these relations numerically. Then, we evaluate the five-point conformal blocks (and their derivatives) using a Pad\' e approximant in order to accelerate the convergence of the series. With this method, we are able to efficiently compute five-point conformal blocks for exchanged operators of fairly high spin with very high accuracy.

Next, we studied the $\langle\sigma\sigma\epsilon\sigma\sigma\rangle$ correlator in the 3d critical Ising model using the truncation technique developed in \cite{Gliozzi:2013ysa, Gliozzi:2014jsa}, and we approximated the truncated part of the correlator with the appropriate contributions from a disconnected five-point correlator. In doing so, we observed a nonnegligible shift in the values of the computed OPE coefficients from those reported in \cite{Poland:2023vpn}. Presumably, the error bars reported there were somewhat underestimated, but it appears that they were of the right order of magnitude. We also note that the error bars get smaller when we add the contributions from the disconnected correlator, which implies that the crossing relation is better satisfied and there is less sensitivity to the choice of derivatives of the crossing relation that we take. Still, our estimated errors  for the OPE coefficients $\lambda_{C\epsilon C}^{n_{IJ}=0,1,2,3}$ and $\lambda_{\epsilon' \epsilon C}$ are one or two orders of magnitude larger than that of the OPE coefficients $\lambda_{T\epsilon T}^{0}$, $\lambda_{T\epsilon C}^{0}$, and $\lambda_{C\epsilon C}^{4}$, which suggests that these OPE coefficients are not being constrained as effectively and may be more sensitive to the error that we make by using the disconnected approximation. It will be interesting to understand how to compute these coefficients more precisely.

One can think of the approximation of truncated operators by the contributions from the disconnected correlator as the leading-order term in the large-spin expansion arising from $(\mathbf{1}, \epsilon)$ exchange. Adding perturbative corrections at large spin (or using the Lorentzian inversion formula) could improve the approximation and consequently the accuracy of the results presented in this paper. A systematic study of the large-spin expansion applied to five-point correlators in the 3d critical Ising model is left for future work.

Another important question for the future is whether the disconnected approximation works well for other correlators, for example $\langle \sigma \sigma \epsilon' \sigma \sigma \rangle$, and for other theories. More generally, this analysis raises the question: what is the criterion for the validity of the disconnected approximation for the truncated part of the correlator? More precisely, can it be connected more quantitatively to the size of the anomalous dimension(s) of the leading Regge trajectories? Since it describes the leading correlator in any holographic theory with a bulk 3-point interaction, we expect that it will work well in many large $N$ theories. However, the calculations in this paper point to the approximation working well even in theories with small $N$. If this is indeed the case, we expect that it will help us to compute many more unknown OPE coefficients in both the critical Ising model and in other interesting CFTs. 

\subsection*{Acknowledgments}
We thank Ilija Buri\' c, Vasiliy Dommes, Rajeev Erramilli, Yin-Chen He, Murat Kolo\u{g}lu, Petr Kravchuk, Matthew Mitchell, Slava Rychkov, Witold Skiba, Lorenzo Quintavalle, David Simmons-Duffin, Yuan Xin, and Zheng Zhou for discussions. The work of D.P. and P.T. is supported by U.S. DOE grant DE-SC00-17660 and Simons Foundation grant 488651 (Simons Collaboration on the Nonperturbative Bootstrap). The work of V.P. is supported by the Perimeter Institute for Theoretical Physics. Computations in this work were performed on the Yale Grace computing cluster, supported by the facilities and staff of the Yale University Faculty of Sciences High Performance Computing Center. This work was performed in part at Aspen Center for Physics, which is supported by National Science Foundation grant PHY-2210452.

\newpage
\appendix

\section{Recursion relation for $c$-coefficients}\label{recrelapp}

In the coordinates $(R, r, \eta_1, \eta_2, \hat{w})$, defined by eq.~(\ref{angluarcoord}), the two quadratic Casimir equations have the schematic form
\begin{equation}
\left(\sum_{i}\mathfrak{R}_{i}(R, r, \eta_1, \eta_2, \hat{w}, \Delta_{12}, \Delta_{3}, \Delta_{45}, \Delta, \Delta', l, l', d)\partial_{i}^{p \leqslant 2} \right)G^{(n_{IJ})}_{(\Delta,l,\Delta',l')} = 0\,,
\end{equation}
where $\partial_{i}^{p \leqslant 2}$ denotes partial derivatives with respect to the cross-ratios $(r_1, \eta_1, r_2, \eta_2, \hat{w})$ whose order is less than or equal to two. The functions $\mathfrak{R}_{i}$ are rational functions of their variables.

We write
\begin{equation}\label{Gaf}
\begin{split}
G^{(n_{IJ})}_{(\Delta,l,\Delta',l')}(R, r, \eta_1, \eta_2, \hat{w}) =& R^{\Delta+\Delta'}r^{\Delta-\Delta'}\times\\
&\sum_{n,m,j_i,k} c\left(\frac{n+m}{2},\frac{n-m}{2},j_1,j_2,k\right)f_{n,m,j_1,j_2,k}(R,r,\eta_1,\eta_2,\hat{w})\,,
\end{split}
\end{equation}
where
\begin{equation}
f_{n,m,j_1,j_2,k}(R,r,\eta_1,\eta_2,\hat{w})=R^{n}r^{m}\eta_{1}^{j_1-k} \eta_{2}^{j_2-k}\hat{w}^{k}\,.
\end{equation}
Then, we substitute eq.~(\ref{Gaf}) in the quadratic Casimir equations and use the following relations satisfied by the functions  $f_{n,m,j_1,j_2,k}$:\footnote{We suppress the $(R,r,\eta_1,\eta_2,\hat{w})$ dependence.}
\begin{equation}
\begin{split}
\frac{\partial}{\partial R}  f_{n,m,j_1,j_2,k} =\, &n f_{n-1,m,j_1,j_2,k}\,,\\
\frac{\partial}{\partial r}  f_{n,m,j_1,j_2,k} =\, &m f_{n,m-1,j_1,j_2,k}\,,\\
\frac{\partial}{\partial \hat{w}}  f_{n,m,j_1,j_2,k} =\, &k f_{n,m,j_1-1,j_2-1,k-1}\,,\\
\frac{\partial}{\partial \eta_1}  f_{n,m,j_1,j_2,k} =\, &(j_1-k)f_{n,m,j_1-1,j_2,k}\,,\\
\frac{\partial}{\partial \eta_2}  f_{n,m,j_1,j_2,k} =\, &(j_2-k)f_{n,m,j_1,j_2-1,k}\,,\\
\end{split}
\end{equation}
and
\begin{equation}
\begin{split}
R f_{n,m,j_1,j_2,k} =\, & f_{n+1,m,j_1,j_2,k}\,,\\
r f_{n,m,,j_1,j_2,k} =\, & f_{n,m+1,j_1,j_2,k}\,,\\
\hat{w} f_{n,m,j_1,j_2,k} =\, & f_{n,m,j_1+1,j_2+1,k+1}\,,\\
\eta_1 f_{n,m,j_1,j_2,k} =\, & f_{n,m,j_1+1,j_2,k}\,, \\
\eta_2 f_{n,m,j_1,j_2,k} =\, & f_{n,m,j_1,j_2+1,k}\,. \\
\end{split}
\end{equation}
By repeatedly applying these relations to the quadratic Casimir equations we first remove all explicit cross-ratio dependence from them. We then isolate the term that multiplies the monomial $f_{n,m,j_1,j_2,k}$, and demand that it vanish identically, to guarantee that the Casimir equations be satisfied, as these functions are linearly independent. This condition can be written as
\begin{equation}\label{recrel}
\sum_{\{\hat{m}_{1},\hat{m}_{2}, \hat{j}_1,\hat{j}_2, \hat{k}\}\in \mathcal{S}_{j}} q_{j}(\hat{m}_1, \hat{m}_2, \hat{j}_1, \hat{j}_2, \hat{k}) c\left(\frac{n+m}{2}+\hat{m}_1, \frac{n-m}{2}+\hat{m}_2, j_1+\hat{j}_1, j_2+\hat{j}_2, k+\hat{k}\right) =0\,,
\end{equation}
where $j$ denotes the Casimir equation we are solving and runs over $j=1,2$. The sets $\mathcal{S}_j$ contain terms such as $(0,0,0,0,0)$, $(0,0,0,0,-1)$, $(0,0,0,-1,-1)$, etc. Each of the sets $\mathcal{S}_{j}$  contains 499 elements. Here, the $q_{j}$ are known rational functions of $\Delta_{12}$, $\Delta_{3}$, $\Delta_{45}$, $\Delta$, $\Delta'$, $l$, $l'$, $d$, and the coefficients $n$, $m$, $j_{1}$, $j_2$ and $k$. Now, instead of solving the Casimir equations perturbatively in $R$ and $r$, we can simply solve~(\ref{recrel}) to compute the $c$-coefficients. The recursion relations and the algorithm for solving them are implemented in the attached {\fontfamily{lmss}\selectfont Mathematica} notebook.

\section{Derivatives and parametrization}

\subsection{Sets of derivatives used in the four-point bootstrap}

Here, we present the sets of derivatives we use to obtain constraints when bootstrapping the four-point correlator that give $\Delta_{\sigma}$ closest to its known value when the mean-field theory contributions are not included.

\begin{equation}\label{fourpsetv1}
\begin{split}
\mathcal{D}=\{&1, \partial_{b}, \partial_{a}^2, \partial_{b}^2, \partial_{b}^3, \partial_{a}^4  \}\,,\\
\mathcal{D}=\{&1, \partial_{b}, \partial_{a}^2, \partial_{b}\partial_{a}^2, \partial_{a}^4, \partial_{a}^2\partial_{b}^2, \partial_{b}\partial_{a}^4\}\,,\\
\mathcal{D}=\{&1, \partial_{b}, \partial_{a}^2, \partial_{b}^2, \partial_{b}\partial_{a}^2, \partial_{a}^4, \partial_{a}^2\partial_{b}^2, \partial_{b}\partial_{a}^4\}\,,\\
\mathcal{D}=\{&1, \partial_{b}, \partial_{a}^2, \partial_{b}^2, \partial_{b}\partial_{a}^2, \partial_{b}^3, \partial_{a}^4, \partial_{a}^2\partial_{b}^2, \partial_{b}\partial_{a}^4\}\,.
\end{split}
\end{equation}

Next, we write the list of derivatives we use to compute constraints when bootstrapping the four-point correlator that give $\Delta_{\sigma}$ closest to its known value when the mean-field theory contributions are included.

\begin{equation}\label{fourpsetsv2}
\begin{split}
\mathcal{D}=\{&1, \partial_{b}, \partial_{a}^2, \partial_{b}^2, \partial_{b}^3, \partial_{a}^2\partial_{b}^3  \}\,,\\
\mathcal{D}=\{&1, \partial_{b}, \partial_{a}^2, \partial_{b}^2, \partial_{b}\partial_{a}^2, \partial_{b}^3, \partial_{a}^2\partial_{b}^3  \}\,,\\
\mathcal{D}=\{&1, \partial_{b}, \partial_{b}^2, \partial_{b}\partial_{a}^2, \partial_{b}^3, \partial_{b}^2\partial_{a}^2, \partial_{b}^4, \partial_{b}^3\partial_{a}^2  \}\,,\\
\mathcal{D}=\{&1, \partial_{b}, \partial_{a}^2, \partial_{b}^2, \partial_{b}\partial_{a}^2, \partial_{b}^3, \partial_{b}^2\partial_{a}^2, \partial_{b}^4, \partial_{b}^3\partial_{a}^2  \}\,.
\end{split}
\end{equation}

\subsection{Parametrization of the cross-ratios in the five-point bootstrap}\label{parametrization}

For convenience, we use the following parametrization of the cross-ratios in the five-point correlator:
\begin{equation}\label{new-coordinates}
\begin{split}
u_1'=\,& \frac{1}{4} \left((a^{-}+a^{+})^2-b^{-}-b^{+}\right) \,,\\
v_1'=\,& \frac{1}{4} \left((a^{-}+a^{+}-2)^2-b^{-}-b^{+}\right)\,,\\
u_2'=\,& \frac{1}{4} \left((a^{+}-a^{-})^2+b^{-}-b^{+}\right) \,,\\
v_2'=\,& \frac{1}{4} \left((a^{+}-a^{-}-2)^2+b^{-}-b^{+}\right)\,,\\
w' =\,& \frac{1}{4} \Big((a^{-}+a^{+})^2+2 (a^{+}-a^{-}-2) (a^{-}+a^{+})+(a^{+}-a^{-}-4) (a^{+}-a^{-})\\
&+2 (2 w-1) \sqrt{b^{+}-b^{-}} \sqrt{b^{-}+b^{+}}-2 b^{+}+4\Big)\,.
\end{split}
\end{equation}
In these coordinates, the configuration (\ref{configuration}) is given by
\begin{equation}\label{configurationab}
a^{+}=1\,,\qquad b^{+}=-3\,, \qquad a^{-}=b^{-}=0\,, \qquad w=\frac{1}{2}\,.
\end{equation}

We take derivatives $\mathcal{D}_i$ of the crossing relation with respect to the $(a^+, b^+, a^-, b^-, w)$ coordinates and we evaluate the derivatives at eq.~(\ref{configurationab}).

\subsection{Sets of derivatives used in the five-point bootstrap}

Here, we present the list of derivatives we use to calculate constraints that give $\Delta_{\sigma}$ closest to its known value when the set $\mathcal{S}$ is given by eq.~(\ref{sdef}).

\begin{equation}\label{setsv1}
\begin{split}
\mathcal{D}=\{&1,\partial_{w},\partial_{b^{+}}, \partial_{w}\partial_{b^{+}}, \partial_{a^{+}}^2\partial_{b^{+}}, \partial_{b^+}^2, \partial_{b^+}^3, \partial_{b^-}^2, \partial_{a^-}^2\partial_{w}, \partial_{b^-}^2\partial_{b^+}  \}\,,\\
\mathcal{D}=\{&1, \partial_{w},\partial_{b^{+}}, \partial_{w}\partial_{b^{+}}, \partial_{b^{+}}\partial_{w}^2, \partial_{a^+}^2, \partial_{b^+}^2, \partial_{b^+}^2\partial_{w}, \partial_{a^-}^2, \partial_{a^-}^2\partial_{w}, \partial_{a^-}^2\partial_{b^+} \}\,,\\
\mathcal{D}=\{&1, \partial_{w},\partial_{b^{+}}, \partial_{w}\partial_{b^{+}}, \partial_{b^{+}}\partial_{w}^2, \partial_{a^+}^2, \partial_{b^+}^2, \partial_{b^+}^2\partial_{w}, \partial_{a^-}^2, \partial_{a^-}^2\partial_{w}, \partial_{b^{-}}^2\partial_{b^+},\partial_{a^-}^2\partial_{b^+} \}\,,\\
\mathcal{D}=\{&1, \partial_{w},\partial_{b^{+}}, \partial_{w}\partial_{b^{+}}, \partial_{b^{+}}\partial_{w}^2, \partial_{a^+}^2, \partial_{b^+}^2, \partial_{b^+}^2\partial_{w}, \partial_{b^-}^2, \partial_{a^-}^2, \partial_{a^-}^2\partial_{w}, \partial_{b^{-}}^2\partial_{b^+},\partial_{a^-}^2\partial_{b^+} \}\,,\\
\mathcal{D}=\{&1, \partial_{w},\partial_{b^{+}}, \partial_{w}\partial_{b^{+}}, \partial_{b^{+}}\partial_{w}^2, \partial_{a^+}^2, \partial_{b^+}^2, \partial_{b^+}^2\partial_{w}, \partial_{b^-}^2, \partial_{a^-}^2, \partial_{a^-}^2\partial_{w}, \partial_{b^{-}}^2\partial_{b^+},\partial_{a^-}^2\partial_{b^+}, \partial_{b^-}^2\partial_{w} \}\,,\\
\mathcal{D}=\{&1, \partial_{w}, \partial_{w}^2, \partial_{b^+}, \partial_{b^+}\partial_{w}, \partial_{b^+}\partial_{w}^2, \partial_{a^+}^2, \partial_{a^+}^2\partial_{b^+}, \partial_{b^+}^2, \partial_{b^+}^2\partial_{w}, \partial_{a^-}^2, \partial_{b^-}^2\partial_{w}, \partial_{a^-}^2\partial_{w}, \partial_{b^-}^2\partial_{b^+}, \\
&\partial_{a^-}^2\partial_{b^+} \}\,,\\
\mathcal{D}=\{&1, \partial_{w}, \partial_{w}^2, \partial_{b^+}, \partial_{b^+}\partial_{w}, \partial_{b^+}\partial_{w}^2, \partial_{a^+}^2, \partial_{a^+}^2\partial_{w}, \partial_{b^+}^2, \partial_{b^+}^3, \partial_{b^-}^2, \partial_{a^-}^2, \partial_{b^-}^2\partial_{w}, \partial_{a^-}^2\partial_{w}\,,\\
& \partial_{b^-}^2\partial_{b^+}, \partial_{a^-}^2\partial_{b^+} \}\,,\\
\mathcal{D}=\{&1, \partial_{w}, \partial_{w}^2, \partial_{w}^3, \partial_{b^+}, \partial_{b^+}\partial_{w}, \partial_{b^+}\partial_{w}^2, \partial_{a^+}^2, \partial_{a^+}^2\partial_{b^+}, \partial_{b^+}^2, \partial_{b^+}^3, \partial_{b^-}^2, \partial_{a^-}^2, \partial_{b^-}^2\partial_{w}\,,\\
& \partial_{a^-}^2\partial_{w}, \partial_{b^-}^2\partial_{b^+}, \partial_{a^-}^2\partial_{b^+}  \}\,.
\end{split}
\end{equation}

Next, we write the list of derivatives we use to calculate constraints that give $\Delta_{\sigma}$ closest to its known value when the set $\mathcal{S}$ is given by eq.~(\ref{sdef}), with the contribution $(\epsilon, S_{\mu\nu\rho\sigma\alpha\beta})$ included.
\begin{equation}\label{setsv2}
\begin{split}
\mathcal{D}=\{&1, \partial_{b^+}, \partial_{b^+}\partial_{w}, \partial_{b^+}\partial_{w}^2, \partial_{a^+}^2, \partial_{b^+}^2, \partial_{b^+}^2\partial_{w}, \partial_{b^-}^2, \partial_{a^-}^2, \partial_{a^-}^2\partial_{b^+}  \}\,,\\
\mathcal{D}=\{&1, \partial_{b^+}, \partial_{b^+}\partial_{w}, \partial_{b^+}\partial_{w}^2, \partial_{a^+}^2, \partial_{b^+}^2, \partial_{b^+}^2\partial_{w}, \partial_{b^-}^2, \partial_{a^-}^2, \partial_{b^-}^2\partial_{b^+}, \partial_{a^-}^2\partial_{b^+}  \}\,,\\
\mathcal{D}=\{&1, \partial_{w}, \partial_{b^+}, \partial_{b^+}\partial_{w}, \partial_{b^+}\partial_{w}^2, \partial_{a^+}^2, \partial_{b^+}^2, \partial_{b^+}^2\partial_{w}, \partial_{b^-}^2, \partial_{a^-}^2, \partial_{a^-}^2\partial_{w}, \partial_{a^-}^2\partial_{b^+} \}\,,\\
\mathcal{D}=\{&1, \partial_{w}, \partial_{b^+}, \partial_{b^+}\partial_{w}, \partial_{b^+}\partial_{w}^2, \partial_{a^+}^2, \partial_{b^+}^2, \partial_{b^+}^2\partial_{w}, \partial_{a^-}^2, \partial_{b^-}^2\partial_{w}, \partial_{a^-}^2\partial_{w}, \partial_{b^-}^2\partial_{b^+}, \partial_{a^-}^2\partial_{b^+} \}\,,\\
\mathcal{D}=\{&1, \partial_{w}, \partial_{b^+}, \partial_{b^+}\partial_{w}, \partial_{b^+}\partial_{w}^2, \partial_{a^+}^2, \partial_{b^+}^2, \partial_{b^+}^2\partial_{w}, \partial_{b^-}^2, \partial_{a^-}^2, \partial_{b^-}^2\partial_{w}, \partial_{a^-}^2\partial_{w}, \partial_{b^-}^2\partial_{b^+}, \partial_{a^-}^2\partial_{b^+} \}\,,\\
\mathcal{D}=\{&1, \partial_{w}, \partial_{b^+}, \partial_{b^+}\partial_{w}, \partial_{b^+}\partial_{w}^2, \partial_{a^+}^2, \partial_{a^+}^2\partial_{b^+}, \partial_{b^+}^2, \partial_{b^+}^2\partial_{w}, \partial_{b^-}^2, \partial_{a^-}^2, \partial_{b^-}^2\partial_{w}, \partial_{a^-}^2\partial_{w}, \partial_{b^-}^2\partial_{b^+}\,,\\
& \partial_{a^-}^2\partial_{b^+} \}\,,\\
\mathcal{D}=\{&1, \partial_{w}, \partial_{w}^2, \partial_{b^+}, \partial_{b^+}\partial_{w}, \partial_{b^+}\partial_{w}^2, \partial_{a^+}^2, \partial_{a^+}^2\partial_{b^+}, \partial_{b^+}^2, \partial_{b^+}^2\partial_{w}, \partial_{b^-}^2, \partial_{a^-}^2, \partial_{b^-}^2\partial_{w}, \partial_{a^-}^2\partial_{w}\,,\\
& \partial_{b^-}^2\partial_{b^+}, \partial_{a^-}^2\partial_{b^+} \}\,,\\
\mathcal{D}=\{&1, \partial_{w}, \partial_{w}^2, \partial_{w}^3, \partial_{b^+}, \partial_{b^+}\partial_{w}, \partial_{b^+}\partial_{w}^2, \partial_{a^+}^2, \partial_{a^+}^2\partial_{w}, \partial_{b^+}^2, \partial_{b^+}^3, \partial_{b^-}^2, \partial_{a^-}^2, \partial_{b^-}^2\partial_{w}\,,\\
& \partial_{a^-}^2\partial_{w}, \partial_{b^-}^2\partial_{b^+}, \partial_{a^-}^2\partial_{b^+} \}\,.
\end{split}
\end{equation}

Next, we write the list of derivatives we use to compute constraints that give $\Delta_{\sigma}$ closest to its known value when the set $\mathcal{S}$ is given by eq.~(\ref{sdef}), with the contributions $(\epsilon, S_{\mu\nu\rho\sigma\alpha\beta})$ and $(\epsilon, \mathcal{E}_{\mu_1\ldots\mu_8})$ included.

\begin{equation}\label{setsv3}
\begin{split}
\mathcal{D}=\{&1, \partial_{b^+}, \partial_{b^+}\partial_{w}, \partial_{b^+} \partial_{w}^2, \partial_{b^+}^2\partial_{w}, \partial_{b^-}^2, \partial_{a^-}^2, \partial_{b^-}^2\partial_{w}, \partial_{a^-}^2\partial_{w}, \partial_{b^-}^2\partial_{b^+} \}\,,\\
\mathcal{D}=\{&1, \partial_{b^+}, \partial_{b^+}\partial_{w}, \partial_{b^+} \partial_{w}^2, \partial_{a^+}^2, \partial_{b^+}^2, \partial_{b^+}^2 \partial_{w}, \partial_{b^-}^2, \partial_{a^-}^2, \partial_{b^-}^2\partial_{b^+}, \partial_{a^-}^2 \partial_{b^+} \}\,,\\
\mathcal{D}=\{&1, \partial_{w}, \partial_{b^+}, \partial_{b^+}\partial_{w}, \partial_{b^+} \partial_{w}^2, \partial_{a^+}^2, \partial_{b^+}^2, \partial_{b^+}^2 \partial_{w}, \partial_{b^-}^2, \partial_{a^-}^2, \partial_{a^-}^2\partial_{w}, \partial_{a^-}^2 \partial_{b^+} \}\,,\\
\mathcal{D}=\{&1, \partial_{w}, \partial_{b^+}, \partial_{b^+}\partial_{w}, \partial_{b^+} \partial_{w}^2, \partial_{a^+}^2, \partial_{b^+}^2, \partial_{b^+}^2 \partial_{w}, \partial_{b^-}^2, \partial_{a^-}^2, \partial_{b^-}^2\partial_{w}, \partial_{a^-}^2\partial_{w}, \partial_{a^-}^2 \partial_{b^+} \}\,,\\
\mathcal{D}=\{&1, \partial_{w}, \partial_{b^+}, \partial_{b^+}\partial_{w}, \partial_{b^+} \partial_{w}^2, \partial_{a^+}^2, \partial_{a^+}^2\partial_{b^+}, \partial_{b^+}^2, \partial_{b^+}^2\partial_{w}, \partial_{a^-}^2, \partial_{b^-}^2\partial_{w}, \partial_{a^-}^2\partial_{w}, \partial_{b^-}^2\partial_{b^+}\,,\\
& \partial_{a^-}^2\partial_{b^+}  \}\,,\\
\mathcal{D}=\{&1, \partial_{w}, \partial_{b^+}, \partial_{b^+}\partial_{w}, \partial_{b^+} \partial_{w}^2, \partial_{a^+}^2, \partial_{a^+}^2\partial_{b^+}, \partial_{b^+}^2, \partial_{b^+}^2\partial_{w}, \partial_{b^-}^2, \partial_{a^-}^2, \partial_{b^-}^2\partial_{w}, \partial_{a^-}^2\partial_{w}, \partial_{b^-}^2\partial_{b^+}\,,\\
& \partial_{a^-}^2\partial_{b^+}  \}\,,\\
\mathcal{D}=\{&1, \partial_{w}, \partial_{w}^2, \partial_{b^+}, \partial_{b^+}\partial_{w}, \partial_{b^+} \partial_{w}^2, \partial_{a^+}^2, \partial_{a^+}^2\partial_{b^+}, \partial_{b^+}^2, \partial_{b^+}^2\partial_{w}, \partial_{b^-}^2, \partial_{a^-}^2, \partial_{b^-}^2\partial_{w}, \partial_{a^-}^2\partial_{w},
\\
& \partial_{b^-}^2\partial_{b^+}, \partial_{a^-}^2\partial_{b^+}  \}\,,\\
\mathcal{D}=\{&1, \partial_{w}, \partial_{w}^2, \partial_{w}^3, \partial_{b^+}, \partial_{b^+}\partial_{w}, \partial_{b^+} \partial_{w}^2, \partial_{a^+}^2, \partial_{a^+}^2\partial_{b^+}, \partial_{b^+}^2, \partial_{b^+}^2\partial_{w}, \partial_{b^-}^2, \partial_{a^-}^2, \partial_{b^-}^2\partial_{w}\,,\\
& \partial_{a^-}^2\partial_{w}, \partial_{b^-}^2\partial_{b^+}, \partial_{a^-}^2\partial_{b^+}  \}\,.
\end{split}
\end{equation}

\newpage
\bibliographystyle{JHEP}
\bibliography{refs-recrel.bib}{}
\end{document}